# Electron-Phonon Coupling, Thermal Expansion Coefficient, Resonance Effect and Phonon Dynamics in High Quality CVD Grown Mono and Bilayer MoSe$_2$

Deepu Kumar[1#], Vivek Kumar[1], Rahul Kumar[2], Mahesh Kumar[2], Pradeep Kumar[1*]

[1]School of Basic Sciences, Indian Institute of Technology Mandi, 175005, India

[2]Department of Electrical Engineering, Indian Institute of Technology Jodhpur, 342001, India

## Abstract

Probing phonons, quasi-particle excitations and their coupling has enriched our understanding of these 2D materials and proved to be crucial for developing their potential applications. Here, we report comprehensive temperature, 4-330 K, and polarization-dependent Raman measurements on mono and bilayer MoSe$_2$. Phonon's modes up to fourth-order are observed including forbidden Raman and IR modes, understood considering Fröhlich mechanism of exciton-phonon coupling. Most notably, anomalous variations in the phonon linewidths with temperature pointed at the significant role of electron-phonon coupling in these systems, especially for the out-of-plane ($A_{1g}$) and shear mode ($E_{2g}^2$), which is found to be more prominent in the narrow-gaped bilayer than the large gapped monolayer. Via polarization-dependent measurements, we deciphered the ambiguity in symmetry assignments, especially to the peaks around ~ 170 cm$^{-1}$ and ~ 350 cm$^{-1}$. Temperature-dependent thermal expansion coefficient, an important parameter for the device performance, is carefully extracted for both mono and bilayer by monitoring the temperature-dependence of the real-part of the phonon self-energy parameter. Our temperature-dependent in-depth Raman studies provide a pave for uncovering the deeper role of phonons in these 2D layered materials from a fundamental as well as application point of view.

[#]E-mail: deepu7727@gmail.com
[*]E-mail: pkumar@iitmandi.ac.in

## 1. Introduction

Group VI transition metal dichalcogenides (TMDCs) in their naturally occurring bulk form have been studied for decades due to their rich physics and industrial applications [1-2]. Research interests in these TMDCs materials have gained a lot of attention after the successful isolation of single-layer graphene from graphite in 2004 [3]. These are the layered materials belonging to the family of 2D materials with a common atomic formula $MX_2$, where M is the transition-metal atom (Mo or W), and X is the chalcogen atom (S, Se or Te) [4-6]. The properties of these 2D materials are strongly dependent on the number of layers[7-8]. For example, these show indirect to direct bandgap transition when the thickness is reduced to a monolayer from the bulk [8]. TMDCs with monolayer thickness have attracted considerable attention due to their specific, electronic, optoelectronic, spin and valley properties, making them a promising materials for high-performance future electronic, optoelectronic, spintronics and valleytronics devices [6, 9-10].

$MoSe_2$ is one of the crucial members of the TMDCs family. It has emerged as a promising candidate for future electronic and optoelectronic applications due to the small direct bandgap, high carrier mobility and high on-off ratio larger than $10^6$ [4, 6]. Further, the smaller bandgap, making the $MoSe_2$ an exciting candidate for the IR light-emitting devices. Bulk (monolayer) $MoSe_2$ shows an indirect (direct) bandgap with a value of 1.1 eV (1.54 eV) [11]. The performance of electronic and optical devices based on the $MoSe_2$ and other these kinds of 2D materials will be significantly influenced by the change of thermal properties such as thermal expansion coefficient (TEC) and thermal conductivity of the materials, with temperature. Generally, 2D materials are supported on some substrates for device applications, like $SiO_2/Si$, which leads to induced strain into the system due to the TEC mismatch between the $MX_2$ and substrate. The developed strain into the system significantly impacts the fundamental properties of these 2D materials [12-13]. Therefore, to increase the performance and reliability of electronic and optical devices based on these TMDCs, it becomes pertinent to understand the behaviour of TEC as a function of temperature and induced strain/stress due to TEC mismatch between the $MX_2$ and substrate. Raman spectroscopic technique has been proved to be very useful for probing 2D as well bulk systems and their various aspects such as layer stacking geometry, strain effect, thermal properties, defects, number of layers, and so on [5,7,12-20]. Several authors have employed temperature-dependent Raman scattering to estimate thermal properties like TEC and thermal conductivity via monitoring the behaviour of the phonon modes as a function of temperature [14-16]. Additionally, anharmonicity present in the material affects the dynamics of the charge carriers via controlling the strength of electron-phonon and phonon-phonon interactions, which may also significantly impact the functioning of the devices. Electron-phonon coupling in 2D materials plays a crucial role in controlling ballistic transport, Raman spectra, and dynamics of an excited state. Anharmonicity, resulting from phonon-phonon interactions, and electron-phonon coupling may be understood by monitoring the phonon modes' temperature-dependent behaviour.

Here, we report an in-depth temperature-dependent Raman study on layered $MoSe_2$, grown by chemical vapour deposition (CVD) method, in a wide temperature range of 4 to 330 K. The measurements were done on both monolayer (1L) and bilayer (2L) $MoSe_2$. We exracted the thermal expansion coefficient by monitoring the temperature dependence of the first-order optical phonon modes. So far, such studies on the $MoSe_2$ have not been undertaken to the best of our knowledge. Also, the temperature-dependent behaviour of the low-frequency interlayer modes, forbidden Raman or IR active modes, and especially second and higher-order phonon modes are not explored. In this paper, we have focused on these unexplored aspects of this material. Interestingly, we observed the broadening of the $A_{1g}$ and $E_{2g}^2$ shear phonon modes in the low temperature window attributed to the electron-phonon coupling. We believe that our detailed studies will pave the way for further studies on MoSe$_2$ and other 2D materials in the future in this direction.

## 2. Results and Discussions

### 2.1 Characterizations of MoSe$_2$

Figure 1 (b) shows the optical micrograph of CVD grown triangular-shaped flake of the monolayer MoSe$_2$ on SiO$_2$/Si substrate. Details about synthesis, Raman and photoluminescence (PL) measurements could also be found in the supplementary part. Raman and PL identify flakes of mono and bilayer (2L), and thickness is confirmed by atomic force microscopy (AFM). The average thickness of the flake from the substrate is evaluated to be around 1.0 nm, which is close to the value for the monolayer MoSe$_2$ [4] (see Fig. 1 (c) ). Figure 1 (d) shows the room temperature PL spectrum, we observed a strong PL emission from the corner, located at ~ 1.53 eV, consistent with the excitonic transition band $A_{ex}$ at the K point of the Brillouin zone (BZ) for the monolayer MoSe$_2$ [19, 21]. Surprisingly, moving from corner to center of the flake, the peak position of the PL signal is red-shifted by a value of ~ 8 to10 meV and is located at 1.52 eV at the center of the flake. We also noticed that the intensity of the PL signal in the central region is approximately two times quenched in comparison to that of the corner region.

Figure 1 (e) shows the room temperature Raman spectrum of the MoSe$_2$ flake. We find that the spectra exhibit a strong peak at ~ 240.3 cm$^{-1}$, and a very weak peak at ~ 286.1 cm$^{-1}$ corresponds to the Raman active out-of-plane ($A_{1g}$) and in-plane ($E_{2g}^1$) vibrational modes, respectively. It should be noted that with changes in the number of layers for these 2D systems, symmetry representation of phonon modes changes. We have used the terminology as used for the bulk system for both monolayer and bilayer for convenience. The frequency difference between $A_{1g}$ and $E_{2g}^1$ modes is found to be ~ 45.6 cm$^{-1}$ indicating the monolayer of MoSe$_2$ flake [22]. We did not observe the Raman peak associated with the interlayer interaction ($E_{2g}^2$) mode further suggesting that the flake is a monolayer. AFM surface morphology and the Raman and PL intensity mapping are used to check the uniformity of the MoSe$_2$ flake. AFM surface morphology shows the remarkable uniformity of the MoSe$_2$ flake, see Fig 1(c). Figure 1 (f) illustrates typical PL mapping of the entire triangular flake performed at room temperature, showing the PL intensity distribution of the observed PL signal corresponding to the $A_{ex}$ exciton. From Fig. 1 (f), we could see that the PL intensity distribution is nearly uniform at the edges and corners of the MoSe$_2$ flake, while a significant quenching in intensity is observed at the center of the flake. Figure 1 (g-j) shows the Raman intensity mapping performed at room temperature for the characteristic $E_{1g}$, $A_{1g}$, $E_{2g}^1$ and $A_{2u}^2$ modes from the entire flake of MoSe$_2$, respectively. We observe that the Raman intensity distribution of $A_{1g}$ and $E_{2g}^1$ mode is nearly uniform across the whole flake, indicating the uniformity of the MoSe$_2$ flake. PL mapping yields more information about the uniformity and quality of the samples than does Raman mapping because the PL emission is very sensitive to the thickness and defects of the materials. In addition to the quenching in the PL band at the center, we also notice the broadening (see Fig. 1 (d). The broad PL band at the center may result from defects that reduce the lifetime of excitons, causing the broadening of the PL band. Meanwhile, defects may also offer non-radiative channels, which may also quench the intensity of the PL band at the center.

Figure 2 (a) shows the optical micrograph of CVD grown MoSe$_2$ flake consisting of the two different thickness regions. We have selected two regions of interest from the flake exhibiting layers of different thickness. Area represented by F1 and F2 exhibit monolayer and bilayer MoSe$_2$, respectively. Figure 2(b) depicts the room temperature Raman spectrum collected from the F1 and F2 regions. The first-order Raman active $A_{1g}$ and $E_{2g}^1$ modes are observed at 240.1 cm$^{-1}$, 241.1

cm$^{-1}$; 285.6 cm$^{-1}$, 284.9 cm$^{-1}$ for F1 and F2 regions, respectively. The frequency difference between $E_{2g}^1$ and $A_{1g}$ modes is observed to be 45.5 cm$^{-1}$ and 43.8 cm$^{-1}$ for the F1 and F2 region, suggesting the monolayer and bilayer MoSe$_2$, respectively [22]. The $A_{1g}$ mode softens while $E_{2g}^1$ mode stiffens as we move from bilayer to monolayer region, which is in line with the previous reports [7, 22]. We notice that for the F1 region, no additional mode is observed in the low frequency (< 50) range. But for the F2 region, we observed one additional mode, centered at ~ 17.4 cm$^{-1}$ attributed to the interlayer shear mode ($E_{2g}^2$) for the bilayer MoSe$_2$ [23], which further indicates the F1 and F2 regions exhibit monolayer and bilayer thickness, respectively. Observed Raman modes near ~ 170 cm$^{-1}$ and ~352 cm$^{-1}$ have been assigned as $E_{1g}$ and $A_{2u}^2$ modes, respectively, in earlier studies [19] However, the peak observed near 352 cm$^{-1}$ is also assigned as $B_{2g}^1 / A_{2u}$ [22-23], suggesting the ambiguity in assigning proper symmetry to this mode. $E_{1g}$ mode is Raman active but is normally forbidden in backscattering Raman scattering measurements. While $A_{2u}^2$ mode is infrared active (IR) and is associated with the out-of-plane vibration of both the Mo and Se atoms, see inset in Fig 1(e). At room temperature, for the case of a monolayer, $E_{1g}$ and $A_{2u}^2$ modes are observed at 170.4 cm$^{-1}$ and 352.8 cm$^{-1}$, respectively. As the thickness changes from monolayer to the bilayer, $E_{1g}$ mode is strongly redshifted and is observed at 167.9 cm$^{-1}$, while $A_{2u}^2$ mode remains unchanged. Moreover, both modes are prominent in bilayer as compared to in the case of the monolayer.

Figure 2 (c) shows the room temperature PL spectrum of the MoSe$_2$ flake collected from both the F1 (black) as well as F2 (red) regions. For the F1 area, we observed a very intense PL signal centered at ~ 1.54 eV and is very close to $A_{ex}$ for the monolayer MoSe$_2$. As we move from F1 to F2 region, PL signal is strongly quenched by ~ 40 times compared to that of F1 region, is centered at ~ 1.48 eV and is redshifted by ~ 60 meV ( see bottom inset in Fig. 2 (c)). The quenching in intensity and red-shift nature of the PL signal for the bilayer (F2 region) may be due to the transition from direct bandgap in the monolayer to indirect bandgap in the bilayer MoSe$_2$. Figures 2 (d-h) show the Raman intensity mapped of the first-order modes $E_{2g}^2$, $E_{1g}$, $A_{1g}$, $E_{2g}^1$ and $A_{2u}^2$, respectively of the entire flake. Raman intensity map of the $A_{1g}$ mode demonstrates that the bilayer MoSe$_2$ regions give a higher Raman intensity than monolayer regions. In contrast, opposite behaviour is observed for the case of $E_{2g}^1$ mode. We also observed that the intensity distribution of the modes is uniform across monolayer and bilayer region in the entire flake, reflecting the uniformity of our CVD grown MoSe$_2$. Moreover, the Raman intensity map of the $E_{1g}$ and $A_{2u}^2$ modes shows that both these modes are significantly intense in bilayer compared to monolayer. To confirm further, we also performed similar measurements on another triangular-shaped MoSe$_2$ flake consisting of both monolayer and bilayer regions; for details, see Fig. S2 in supplementary information and details therein.

## 2.2 Multi-phonons Raman scattering in MoSe$_2$

Figures 3 (a) and 3 (b) show the Raman spectrum of monolayer and bilayer MoSe$_2$ at 4 K, in a spectral range of 10-640 cm$^{-1}$, respectively. In the yellow shaded area, insets show the amplified spectra in the spectral range of 80-200 cm$^{-1}$, 260-500 cm$^{-1}$ and 540-640 cm$^{-1}$. The spectra are fitted with a sum of Lorentzian functions to extract mode frequency ($\omega$), full width at half maximum (FWHM) and intensity of the individual mode. The observed modes are in excellent agreement with the recently published reports on high-quality MoSe$_2$ sample grown by vapour phase chalcogenization and mechanical exfoliation method [19, 24]. For convenience, we have labelled the observed modes as S1 to S20, and the observed modes and their corresponding symmetries are listed in Table-I, for details about the phonon modes at the gamma point (see supplementary part).

The modes symmetry assignment is done according to the previous reports and our polarization-dependent measurements. The interlayer mode $E_{2g}^2$ (S1) is observed at 17.4 cm$^{-1}$ for bilayer (absent for 1L). The first-order modes $A_{1g}$ (S7) and $E_{2g}^1$ (S8) are observed at 242.3 cm$^{-1}$, 242.2 cm$^{-1}$; 287.9 cm$^{-1}$ 285.9 cm$^{-1}$ in monolayer, and bilayer. Forbidden first-order Raman active modes in backscattering geometry $E_{1g}$ (S6) and IR active $A_{2u}^2$ (S12) are observed at 174.1 cm$^{-1}$, 168.8 cm$^{-1}$; 355.1 cm$^{-1}$ 354 cm$^{-1}$ in monolayer and bilayer, respectively. In addition to the well-known first-order optical modes, $E_{2g}^2$, $E_{1g}$, $A_{1g}$, $E_{2g}^1$ and $A_{2u}^2$, we also observe first-order longitudinal acoustic ($LA$), transverse ($TA$) and out-of-plane ($ZA$) modes near $M$ or $K$ symmetry points in the BZ along with a large number of second- or higher-order phonon modes. The first-order $LA$ mode near the $M$ point of the BZ is observed at 152.1 cm$^{-1}$ and 147.3 cm$^{-1}$ in monolayer and bilayer MoSe$_2$, respectively. Towards the low-frequency side of $LA(M)$ (S5) mode, two weak modes S2 and S3, are observed at 124.8 (129.7) cm$^{-1}$, and 121.9 (127.6) cm$^{-1}$ in monolayer (bilayer); and are assigned as $TA$ and $ZA$, respectively, along $M-K$ direction in the BZ [19]. For the case of monolayer, overtones and combinations of optical and acoustical phonon modes from the $M$ symmetry point of the BZ are observed at 140.7 cm$^{-1}$ ($E_{2g}^1 - LA(M)$; S4), 305.5 cm$^{-1}$ ($2LA(M)$; S9), 364.3 cm$^{-1}$ ($A_{1g}(M) + LA(M)$; S13), 414.6 cm$^{-1}$ ($TA(M) + 2LA(M)$; S14), 432.7 cm$^{-1}$ ($E_{2g}^1(M) + LA(M)$; S15), 457.2 cm$^{-1}$ ($3LA(M)$; S17), 569.7 cm$^{-1}$ ($TA(M) + 3LA(M)$; S18), 584.6 cm$^{-1}$ ($E_{2g}^1(M) + 2LA(M)$; S19) and at 598.2 cm$^{-1}$ ($4LA(M)$; S20). The observed second- and higher-order phonon modes and their corresponding symmetry assignment are given in Table-I for the case of the bilayer. Further, we notice a mode S16 at ~ 444.6 (441.7) cm$^{-1}$ in monolayer (bilayer) at 4K, while at room temperature, this mode is observed at ~ 441.5 (339.1) cm$^{-1}$ in monolayer (bilayer) MoSe$_2$. The energy of this mode is close to the sum of $A_{2u}^2$ (S12) and $LA$ (S5) modes from the $M$ point of the BZ. Interestingly, this mode is absent when the spectra are excited using a 632.8 nm laser (see Fig. S3).

To decipher the symmetry assignment and to understand the angle-dependent nature of the phonon modes with respect to the incident photons polarisation direction, especially for $E_{1g}$ (S6) and $A_{2u}^2$ (S12) modes near 170 cm$^{-1}$ and 350 cm$^{-1}$, respectively; we carried out detailed polarised Raman scattering measurements for both monolayer and bilayer of MoSe$_2$. The polarization-dependent measurements were done by rotating the direction of the incident light with an angle ($\theta$) by keeping fixed the position of the sample and direction of the scattered light as described in Ref [25-26]. Insets in the green shaded area, see Fig. 3 (a) and 3 (b), are the angular dependence of the intensity polar plots of the modes $E_{2g}^2$ (S1), $E_{1g}$ (S6), $A_{1g}$ (S7), $E_{2g}^1$ (S8) and $A_{2u}^2$ (S12) for both monolayer and bilayer. The intensity of the $E_{2g}^2$ (S1) mode shows isotropic nature with respect to the polarisation angle, i.e. intensity is invariant with respect to the rotation of polarisation angle. However, the intensity of the $A_{1g}$ mode shows two-fold symmetric nature, i.e. it has a maximum intensity at both $0^0$ and $180^0$, while the intensity approaches zero at $90^0$ and $270^0$. The angular dependence of the intensity of the mode $E_{1g}$ (S6) is similar to $E_{2g}^2$ mode (i.e. intensity is independent of polarisation angle). The polarisation-dependent results discussed above could also be seen in the 2D colour contour maps of the Raman intensity versus Raman shift and as a function of polarisation angle, which are shown as insets for (a) $A_{1g}$ and (b) $E_{2g}^2$ and $E_{1g}$ in Fig. 3 in the grey colour area for monolayer and bilayer MoSe$_2$, respectively. The intensity of the $E_{2g}^1$ (S8) mode showed slight dependence on the rotation angle. Its intensity is little more at $0^0$ than that at $90^0$, while angular dependence of the intensity of the mode $A_{2u}^2$ (S12) is similar to that of $A_{1g}$ mode. Based on our polarization-dependent Raman observations and IR reflectance observations from literature [2], we attribute modes near ~ 170 cm$^{-1}$ (S6) and ~ 350 cm$^{-1}$ (S12) to the first-order Raman active $E_{1g}$ mode and IR active $A_{2u}^2$ mode, respectively.

The observed variation in intensity as a function of polarization angle may be understood within a semi-classical approximation. As the incident and scattered polarised light lie in the XY plane, the unit vector associated with incident ($\hat{e}_i$) and scattered ($\hat{e}_s$) light of polarisation may be decomposed as [$cos\,(\theta+\theta_0)$, $sin\,(\theta+\theta_0)$, 0)] and [$cos\,(\theta_0)$, $sin\,(\theta_0)$, 0)], respectively, where $\theta_0$ is an arbitrary angle from the $x$-axis and $\theta$ varies from $0^0$ to $360^0$. Within the semi-classical approximation, Raman scattering intensity of the first-order phonon modes is given as $I_{int} = |\hat{e}_s^t.R.\hat{e}_i|^2$, where R is the Raman tensor [27-28]. Using the above expression and Raman tensor [28], the intensity of the $E_{1g}$, $A_{1g}$ and $E_{2g}^1$ modes for our experimental geometry is given as $I_{E_{1g}} = 0$, $I_{A_{1g}} = a\cos^2\theta$ and $I_{E_{2g}^1} = d^2(\cos^2\theta + \sin^2\theta)$, respectively. Within the semi-classical approximation, the following observations can be made: (i) Intensity of the $E_{1g}$ mode is zero, suggesting it should be absent in the backscattering geometry. (ii) Intensity of the $A_{1g}$ mode is maximum when $\hat{e}_i$ and $\hat{e}_s$ are parallel to each other, i.e. $\theta = 0^0$, and it reduces to zero when $\theta = 90^0$. (iii) Intensity of the $E_{2g}^1$ mode remains invariant with respect to rotation of polarisation angle. From our above discussions, we may conclude that within the semi-classical approximation, the modes with $E$ type symmetry are not affected by polarisation configuration and is either observed or forbidden in both parallel and cross-polarization configuration. However, modes with $A$ type symmetry are strongly affected by the polarisation configuration and can be observed only in parallel polarisation configuration. Except for $E_{2g}^1$ (S8), the solid lines are the fitted curves from the above expressions suggesting that the experimental results are in very good agreement with the semi-classical approximation. The intensity pattern of the mode $E_{2g}^1$ in both monolayer and bilayer MoSe$_2$, is slightly smaller in cross-polarization as compared to parallel polarisation configurations, and forming a semi-lobe kind of structure (see Fig. 3 (a) and 3 (b)). Intensity of the $E_{2g}^1$ mode may be fitted well using the combined functions i.e. $(d^2+a^2)\cos^2\theta + d^2\sin^2\theta$, and we can see that the overall fitting is modest (see Fig. 3). The observed anisotropic nature of $E_{2g}^1$ mode may arise due the strong photon-electron-phonon coupling, which may be understood within the quantum mechanical picture [26].

**2.3 Thermal expansion coefficient and temperature-dependent frequency of the first-order optical modes**

To understand the temperature dependence of the phonon modes quantitatively, we extracted self-energy parameters such as mode frequency ($\omega$) and FWHM ($\Gamma$) of the phonon modes using Lorentzian functions fitting. Figure 4 (a) illustrates temperature dependence of the frequency of the $A_{1g}$ (S7) and $E_{2g}^1$ (S8) phonon modes for monolayer MoSe$_2$. Both these modes, i.e. $A_{1g}$ and $E_{2g}^1$, stiffens with decreasing temperature down to ~ 80 K, and below this, a sudden rise (drop) in frequency is observed for the case of $A_{1g}$ ($E_{2g}^1$). On further cooling, till 4K, both these modes remain nearly temperature independent. Figure 4 (b) illustrates temperature dependence of the frequency of $A_{1g}$ and $E_{2g}^1$ phonon modes for bilayer MoSe$_2$. Both $A_{1g}$ and $E_{2g}^1$ mode stiffens with the decreasing temperature down to ~ 80 K; interestingly, a softening in frequency is observed for both the phonon modes on further cooling. Figure 5 (a) shows the temperature dependence of the frequency of $E_{1g}$ (S6) and $A_{2u}^2$ (S12) modes for monolayer. We observe that variations in the frequency of $E_{1g}$ and $A_{2u}^2$ modes in temperature window 330 to ~ 80 K is normal, and at ~ 80 K, a sudden drop in frequency is observed for the case of $E_{1g}$ mode, and below 80 K, it again starts to harden on further cooling till 4 K; while frequency of the $A_{2u}^2$ mode remains nearly constant below

~ 80 K. Figure 5 (b) shows the temperature dependence of the frequency of the $E_{2g}^2$ (S1), $E_{1g}$ (S6) and $A_{2u}^2$ (S12) modes for bilayer. Temperature-dependent behaviour of the frequency of the mode $A_{2u}^2$ (see Fig. 5(b)) is similar to that of $A_{1g}$ and $E_{2g}^1$ modes for bilayer (see Fig 4 (b)). The temperature-dependent shift in the frequency of the $E_{2g}^2$ and $E_{1g}$ modes is surprisingly fascinating. Since the overall change in frequency of the $E_{2g}^2$ and $E_{1g}$ modes is minor (~ 0.6 cm$^{-1}$) nevertheless, we could see apparent variations in the mode frequency with temperature, and this can be divided into four regions: (i) Modes show hardening in the temperature range of 330-280 K, (ii) and then shows softening in range of 280-200 K, (iii) afterwards it again show hardening from ~200 K to ~80 K, (iv) and with further decrease in temperature, it again shows softening till the lowest recorded temperature (4 K). Details about the temperature dependence of the first-order acoustic and second and higher-order phonon modes are given in supplementary information (please see supplementary section 3).

The temperature-dependent shift in frequency of the phonon modes of the free-standing MoSe$_2$ may be understood via: (i) anharmonic effect, which arises due to change in the self-energy parameter because of phonon-phonon coupling (ii) quasi-harmonic effect which arises due to thermal expansion of the lattice. The change in the phonon mode frequency as a function of temperature considering the above two effects may be given as

$$\Delta\omega(T) = \Delta\omega_{Anh}(T) + \Delta\omega_E(T) \qquad (1)$$

where $\Delta\omega_{Anh}(T)$ and $\Delta\omega_E(T)$ correspond to change in the mode frequency due to phonon-phonon anharmonic effect and thermal expansion of the lattice, respectively. The first term in Eq$^n$. 1 arises due to a change in phonon self-energy because of the anharmonic effect and is given as [29] $\Delta\omega_{Anh}(T) = A(1+\frac{2}{e^x-1})$, where $x = \hbar\omega/2k_BT$ and A is a self-energy constant parameter, representing the contributions from three phonon anharmonic effect. An optical phonon decays into two phonons with equal frequencies and opposite momentum in the three-phonon anharmonic model. Furthermore, with the variations in temperature, the lattice parameter of the MoSe$_2$ would change due to thermal expansion of the lattice parameter, resulting in a variation in the phonon mode frequency as a function of temperature. The contribution of thermal expansion effect to the shift in phonon mode frequency is given as [30] $\Delta\omega_E(T) = \omega_0 \exp[-3\gamma \int_{T_0}^{T} \alpha_{MoSe_2}(T)dT] - \omega_0$, where $\gamma$ is the Gruneisen parameter of a particular mode and $\alpha_{MoSe_2}(T)$ is temperature-dependent TEC of MoSe$_2$. We note that both positive and negative TEC has been reported for MoSe$_2$ [31-33]. In the present case, MoSe$_2$ is not free-standing but supported by SiO$_2$/Si substrate, and SiO$_2$ has negative (positive) TEC at low (high) temperature [34]. Therefore, in addition to the mentioned two effects above (i.e. anharmonic and thermal expansion effects), thermally induced strain results from TEC mismatch between MoSe$_2$ and substrate should also be considered to understand the net change in the phonon mode frequency with temperature. Change in the phonon mode frequency, considering these effects, as a function of temperature is given as [15, 29-30]

$$\Delta\omega(T) = \Delta\omega_{Anh}(T) + \Delta\omega_E(T) + \Delta\omega_s(T) \qquad (2)$$

The last term $\Delta\omega_s(T)$ is the change in the mode frequency corresponding to the strain effect due to TEC mismatch. It can be expressed as [15] $\Delta\omega_S(T) = \beta\varepsilon(T) = \beta \int_{T_0}^{T} [\alpha_{SiO_2}(T) - \alpha_{MoSe_2}(T)]dT$, where $\beta$ is the strain coefficient of a particular mode and $\alpha_{SiO_2}(T)$ is the temperature-dependent TEC of SiO$_2$. To estimate $\alpha_{MoSe_2}(T)$, we have used the value of $\beta$ ($= \frac{\partial\omega}{\partial\varepsilon}$) as $\beta_{A_{1g}} = -3.7\ cm^{-1}/\%$ and $\beta_{E_{2g}^1} = -1.2\ cm^{-1}/\%$ [12]. TEC of SiO$_2$ was taken from [34] and was integrated out while estimating TEC for MoSe$_2$. The product of the mode Grüneisen parameter and thermal expansion coefficient may be expressed by a polynomial of temperature and is given as

$$\gamma \, \alpha_{MoSe_2}(T) = p_0 + p_1 T + p_2 T^2 \qquad (3)$$

where $p_0$, $p_1$ and $p_2$ are the constant parameters, whose values are obtained as a fitting parameter by the best fit to the temperature dependence of the frequency of the modes. To extract the TEC of MoSe$_2$, we have fitted the frequency of $A_{1g}$ and $E_{2g}^1$ mode in the temperature range of 80-330 K for monolayer and bilayer using the Eq$^n$. 2 and Eq$^n$. 3. The above Eq$^n$. 3 can be used to estimate TEC as a function of temperature. We have adopted the theoretically calculated value of the mode Gruneisen parameter (~ 1.7 for $A_{1g}$ and ~ 0.8 for $E_{2g}^1$) to estimate the TEC of MoSe$_2$ [32] in both out-of-plane and in-plane direction. Figures 4 (c) and 4 (e) show the temperature-dependent TEC of the in-plane ($E_{2g}^1$) and out-of-plane ($A_{1g}$) modes for monolayer and bilayer MoSe$_2$, respectively. Our temperature-dependent results show that the TEC of $A_{1g}$ mode for both system (1L and 2L MoSe$_2$) decreases sharply with decreasing temperature starting from 330 to ~200 K; on further cooling, a mild decrease in TEC is observed. While TEC corresponding to the $E_{2g}^1$ mode decreases with decreasing temperature from 330 to ~200 K, and surprisingly below 180 K, an increase is seen with further lowering the temperature (see shaded area). We also estimated the volumetric TEC given as $\alpha_v = 2\alpha_a + \alpha_c$ [32], where $\alpha_a$ and $\alpha_c$ are the linear TEC for in-plane and out-of-plane direction. Figure 4 (d) and 4 (f) shows the volumetric TEC as a function of temperature. Room temperature linear TEC corresponding to the $E_{2g}^1$ and $A_{1g}$ modes together with volumetric TEC are listed in Table-II. The volumetric TEC values at room temperature are found to be $21.4 \times 10^{-6}$ K$^{-1}$ and $23.5 \times 10^{-6}$ K$^{-1}$ for monolayer and bilayer MoSe$_2$, respectively, see Table-II. The estimated volumetric TEC results are in excellent agreement with previously reported values for monolayer

MoSe$_2$ supported by SiO$_2$/Si substrate [36]. Our estimated linear in-plane TEC value for the monolayer is very close to the theoretically calculated TEC for monolayer MoSe$_2$ [33], while it is nearly ten times smaller than that of experimentally reported in-plane TEC for free-standing monolayer MoSe$_2$ [37]. Furthermore, TEC corresponding to the $A_{1g}$ mode is larger than that of $E_{2g}^1$ mode for monolayer, which is in line with earlier reports on other TMDCs [35, 38]. The larger TEC of $A_{1g}$ mode may be understood as: in layered materials, the out-of-plane direction is confined weekly compared to the in-plane direction; therefore, it is easier to deform the out-of-plane direction than the in-plane direction [39]. With the change in thickness from monolayer to the bilayer, we observed decreased TEC of $A_{1g}$ mode, reflecting the increase in mode strength. For bilayer, an increment in TEC of $E_{2g}^1$ mode is observed compared to that of the monolayer, which differs from the earlier report on MoSe$_2$ [37]. Therefore, further theoretical and experimental studies on these 2D materials are required to decipher discrepancies in TEC, especially for in-plane linear TEC.

Now we will focus on the observed anomalies in the mode frequencies at low temperature. The observed kink in frequencies of the modes at low temperature for monolayer, as shown in Fig. 4 (a) and Fig. 5 (a), may be due to induced strain owing to TEC mismatch between MoSe$_2$ and the substrate. The induced strain due to TEC mismatch may affect the weak van der walls forces, leading to the slippage, realignment or change in surface morphology of MoSe$_2$ films on the substrate and forming of wrinkles or ripples into the system, which may affect the frequency of the modes. Similar anomalies are also observed in the case of other MX$_2$ systems [38, 40]. For bilayer, we observed a decrease in the mode frequency (see Fig 4(b) and Fig 5 (b)) below ~ 100 K. A negative TEC has been reported for MoSe$_2$ at low temperature [33], and this may generate the tensile stress into the system at low temperature, which may give rise to the anomalous decrease in the mode frequency for bilayer at low temperature. To confirm the realignment of MoSe$_2$ films, we did AFM characterizations after the performance of temperature-dependent Raman measurements of the same flake as optically shown in Fig. 2 (a). Figure 2 (i) shows the AFM image, and the insets depict the step height profile. The step height profile along the white line shows the height of monolayer MoSe$_2$ from the substrate. We notice that the average thickness of monolayer (F1 region) from the substrate is observed to be ~ 3 nm which is significantly larger than that of the actual thickness of monolayer MoSe$_2$ from substrate reflecting the slippage or realignment of MoSe$_2$. The step height profile along the blue line shows the height of the bilayer along with monolayer MoSe$_2$ from the substrate. The average thickness of bilayer from the substrate is ~ 2 nm, reflecting a very weak deviation from the reported thickness for the bilayer MoSe$_2$ from the substrate, suggesting the weak effect of induced strain due to TEC mismatch between bilayer and substrate. It is in line with the fact that induced strain due to TEC mismatch gradually decreases with the increasing number of layers and becomes negligible for the bulk.

## 2.4 Electron-Phonon coupling and lifetimes of the first-order optical modes

A strong electron-phonon coupling, which limits the electronic mobility of semiconductors, can significantly affect the self-energy parameters of the phonon modes, and this effect may be captured via a detailed temperature-dependent Raman measurement. The temperature-dependent mobility in semiconductor MoSe$_2$ was attributed to the scattering of carriers by optical phonons which corresponds to the fluctuations of the layer thickness [41], implying that the A$_{1g}$, phonon with atomic displacements along c-axis, and $E_{2g}^2$ phonon, interlayer shear mode, modes may be involved in controlling the mobility of the carriers. In this section, we focus our attention on the temperature dependence of the FWHM of the phonon modes and the role of electron-phonon coupling in the temperature-dependent evolution of FWHM. In a high-quality sample, FWHM of the phonon modes at finite temperature may be affected by the contributions of two factors: (i) phonon-phonon coupling ($\gamma_{ph\text{-}ph}$) (ii) electron-phonon interactions ($\gamma_{e\text{-}ph}$). Therefore, considering these factors, the temperature dependence of the FWHM of the phonon modes may be given as [42-43]

$$\gamma(T) = \gamma_{ph\text{-}ph}(T) + \gamma_{e\text{-}ph}(T) \qquad (4)$$

The first term $\gamma_{ph\text{-}ph}(T)$ arises from decaying an optical phonon into two phonons of the same energy and opposite momentum satisfying the energy and momentum conservation rules. In general, FWHM of the phonon mode increases as temperature increases. With increasing temperature, populations of the phonons also increase as the lifetime of the phonons is inversely proportional to the FWHM; as a result, significant broadening (increase) in FWHM is expected with increasing temperature. The contributions from the three-phonon anharmonic effect to the FWHM of phonons as suggested by Klemens may be expressed as [29] $\gamma_{ph-ph}(T) = \gamma_{ph-ph}(0) + C(1 + \frac{2}{e^x - 1})$, where $x = \hbar\omega/2k_B T$ C is a constant parameter and $\gamma_{ph-ph}(0)$ is the FWHM at 0 K, and the term $\gamma_{ph-ph}(T)$ is expected to dominate at high temperature. The second term $\gamma_{e\text{-}ph}(T)$ arises as a result of contributions from the electron-phonon interactions. Temperature-dependent contribution from the electron-phonon coupling interactions to the FWHM may be given as [42-43] $\gamma_{e-ph}(T) = \gamma_{e-ph}(0)[(\frac{1}{e^{-x}+1}) - (\frac{1}{e^x+1})]$, where $x = \hbar\omega/2k_B T$ and $\gamma_{e-ph}(0)$ is the FWHM resulting from the electron-phonon coupling effect at 0 K, and the term $\gamma_{e-ph}(T)$ is expected to dominate at low temperature. The expression for $\gamma_{e-ph}(T)$ represents the difference in occupations of states below and above the Fermi energy level and may be used to understand the temperature-dependent shift in the FWHM of the phonon modes. Occupations of the filled states below Fermi level decreases with an increase in temperature, while empty states above Fermi level are occupied more and may result in narrowing (broadening) of the FWHM with increasing (decreasing) temperature [42]. Here, a renormalization of the phonon modes may be understood due to phonon induced electron-hole pair creations. With increasing temperature, empty states above the Fermi level starts filling up and this blocks the generation of the phonon induced electron-hole pairs and hence affects the phonon self-energy. In particular, it is expected that at low (high) temperature the phonon lifetime will be less (more) and as a result linewidth will be more (less); based on the pure electron-phonon coupling effect one expects that linewidth will be more at low temperature and less at a higher temperature. Also, the bandgap in monolayer MoSe$_2$ is significantly higher than that in the bilayer; therefore, the effect of electron-phonon coupling is expected to be more visible in bilayer owing to the reduced bandgap.

Figures 6 (a) and 6 (b) show the temperature dependence of the FWHM of the $A_{1g}$ and $E_{2g}^1$ modes for monolayer and bilayer MoSe$_2$, respectively. For monolayer, we could see that FWHM of the $A_{1g}$ mode decreases with decreasing temperature showing normal temperature dependence; on the other hand, for $E_{2g}^1$ mode, we did not observe any apparent variations in FWHM with temperature for both mono and bilayer. For bilayer, FWHM of the $A_{1g}$ mode shows normal temperature-dependent behaviour in the temperature range of 330 to ~ 100 K, and surprisingly below 100 K, an increase in FWHM is observed with further cooling till the lowest recorded temperature (4 K). Figure 5 (c) shows the temperature dependence of the FWHM of the $E_{1g}$ and $A_{2u}^2$ modes for monolayer MoSe$_2$. We notice that the FWHM of the $E_{1g}$ mode decreases with decreasing temperature, while the FWHM of the $A_{2u}^2$ mode shows a non-monotonic trend with temperature. Figure 5 (d) shows the temperature dependence of the FWHM of the $E_{2g}^2$, $E_{1g}$ and $A_{2u}^2$ modes for bilayer. FWHM of the $E_{2g}^2$ mode shows normal temperature-dependence; it decreases with decreasing temperature, from room temperature to ~ 120 K. Quite surprisingly, on further lowering the temperature, it starts to increase till 4 K attributed to the strong electron-phonon coupling. We note that similar behaviour is also observed for the shear modes in Graphene [43]. FWHM of the $E_{1g}$ mode shows normal temperature dependence till ~ 100 K, and it increases slightly below 100 K. On the other hand, the FWHM of $A_{2u}^2$ mode shows normal temperature dependence in the entire temperature range.

FWHM of the modes as a function of temperature are fitted using the above Eq$^n$. 4, the solid red lines in Fig. 5 (c and d) and Fig. 6 (a and b) are the fitted curves, and it is in very good agreement with our experimental data. For the monolayer, fitting parameters $\gamma_{ph-ph}(0)$ and $\gamma_{e-ph}(0)$ of the $A_{1g}$ mode are 1.4 cm$^{-1}$ and 0.02 cm$^{-1}$, respectively, suggesting that the dominating factor is the phonon-phonon anharmonic effect. However, for the case of bilayer $\gamma_{ph-ph}(0)$ and $\gamma_{e-ph}(0)$ of the $A_{1g}$ mode obtained from the fitting are 0.6 cm$^{-1}$ and 0.7 cm$^{-1}$, respectively, suggesting a significant role of election-phonon coupling at low temperature, and which may also explain the observed increase in the FWHM with decreasing temperature below 100 K. Overall $\gamma_{e-ph}$ is substantial in comparison to $\gamma_{ph-ph}$ for bilayer system as anticipated earlier. The fitting parameters obtained from the FWHM for other first-order optical phonon modes are summarised in Table-III. Our observation of an anomalous increase in FWHM at low temperature, especially in the bilayer, deviating from normal temperature behaviour, may be understood keeping the finite role of electron-phonon coupling in these systems.

## 2.5 Infrared, Forbidden Raman and higher-order phonon modes

In addition to the first-order phonon modes, we observed a large number of phonon modes from the high symmetry point of BZ. Further, we also observed first-order phonon modes, which are either forbidden in backscattering geometry or IR active. The appearance of these forbidden and IR active phonons from the BZ center and multi-phonon Raman scattering from other parts of the BZ may be understood via resonance effect [22, 44], Fröhlich mechanism of exciton-phonon coupling [45], and cascade theory of inelastic light scattering [46]. The resonance effect occurs when the laser excitation energy is close to the excitonic energy state. As a result, under the resonance condition, in addition to first-order Raman active phonon, first-order Raman active but backscattering forbidden and IR active modes from BZ center as well as first-order acoustic phonon modes from the high symmetry point of the BZ could also be observed. Furthermore, the resonance excitation also gives rise to intense and enriched Raman spectrum, including second and high-order phonon modes from the high symmetry point of the BZ [44]. Recently, Bilgin *et al.* [19] theoretically estimated the energy of C exciton to be 2.33 eV, while experimentally, this peak is reported at ~ 2.5 eV [47]. In our case, excitation wavelength 532 nm (2.33 eV) is near C exciton energy; and could be expected to be a good condition for near resonance Raman scattering effect. Under the resonance condition, the observation of backscattering forbidden ($E_{1g}$) and IR-active ($A_{2u}^2$) phonons may be understood by the Fröhlich mechanism of exciton-phonon coupling. Fröhlich mechanism of the exciton-phonon coupling may give rise to finite intensity of the forbidden phonons proportional to $(aq)^2$, where $a$ is Bohr radius of exciton, and $q$ is wave vector of phonon. When the Bohr radius is much larger than the lattice parameter, then the forbidden phonon modes may appear in Raman spectra under resonance condition. These conditions may be satisfied easily in TMDCs, where the Bohr radius of the exciton is larger than the lattice parameter [48]. As a result, they can couple to the phonons and give into a finite intensity even for the forbidden phonons. It was recently reported that $E_{1g}$ and $A_{2u}^2$ modes are observed only for excitation energies above 2.2 eV, and they both become intense close to the C exciton [24]. To confirm, we excited the spectra with 632.8 nm (1.96 eV) laser; surprisingly, both $E_{1g}$ and $A_{2u}^2$ modes were not seen in the Raman spectrum, see Fig. S3. Therefore, the appearance of the $E_{1g}$ and $A_{2u}^2$ modes in our Raman spectrum of the MoSe$_2$ using 532 nm laser reflects the resonance effect with *C* exciton. Further, it should be noted that we observed intense phonon modes up to fourth-order overtone $4LA(M)$, and the appearance of high-order phonon modes may be understood by considering the cascade process of the Raman scattering of light [46]. In cases when the energy of the incident photons is less than that of the bandgap energy of the material, i.e. ($E_i < E_g$), the intensity of the n$^{th}$-order phonon modes varies as $g^n$, where *g* is the typical electron-phonon coupling constant which is generally much less than unity, *n* is an order of the phonon modes. Therefore, the intensity of higher-order modes decrease extremely fast as *n* increase, and the higher-order modes are generally expected to be very weak or absent in the Raman spectrum. When the energies of both incident and scattered photons are above the bandgap, i.e. $E_i, E_s > E_g$, inelastic scattering of light occurs via cascade process. In such a case, the intensity of modes are independent of electron-phonon coupling and depends mainly on the dispersion curves of the electron and hole bands; as a result, higher-order phonon modes may appear in the Raman spectrum. We note that we have used a 532 nm (2.33 eV) laser as excitation energy which is above the bandgap energy ($E_g$ ~ 1.55 eV) [11]. Therefore above discussed cascade process of Raman scattering may be easily satisfied, which may give rise to intense high-order phonon modes in the Raman spectrum in our case.

## 3. Conclusion

In conclusion, we performed a comprehensive temperature- and polarisation-dependent Raman study on CVD grown MoSe$_2$ supported by SiO$_2$(~ 300 nm)/Si in a wide temperature and broad spectral range of 10-700 cm$^{-1}$. A large number of phonons modes were observed, up to fourth-order as well forbidden Raman and IR modes, understood by considering the resonance effect, Fröhlich mechanism of exciton-phonon coupling and cascade theory of inelastic light scattering.

The thermal expansion coefficient is extracted for both mono and bilayer MoSe$_2$ as a function of temperature, and the effect of induced strain from the underlying substrate is found to be significant for the case of a monolayer. The observed temperature evolution of the linewidth of the $A_{1g}$ and $E_{2g}^2$ mode suggests that electron-phonon processes are involved in addition to the phonon-phonon anharmonicity, and is found to be dominating in the case of the bilayer.

**Acknowledgement:** PK acknowledges the Department of Science and Technology (DST) and IIT Mandi, India, for the financial support.

**References:**

[1]  T. Sekine et al., Solid State Commu. 35, 371 (1980).
[2]  T. Sekine et al., J. Phys. Soc. Jpn. 49, 1069 (1980).
[3]  K. S. Novoselov et al., Science 306, 666 (2004).
[4]  X. Wang et al., ACS Nano 8, 5125 (2014).
[5]  A. A. Puretzky et al., ACS Nano 9, 6333 (2015).
[6]  S. Larentis et al., App. Phy. Lett. 101, 223104 (2012).
[7]  H. Li, et al., Adv. Funct. Mater. 22, 1385 (2012).
[8]  A. Splendiani et al., Nano Lett.10, 1271(2010).
[9]  J. Isberg et al., Nat. Materials 12, 760 (2013).
[10] N. Kumar et al., Nanoscale 6, 12690 (2014).
[11] S. Tongay et al., Nano Lett.12, 5576 (2012).
[12] M. Yagmurcukardes et al., Phys. Rev. B 97, 115427 (2018).
[13] E. Blundo et al., Phys. Rev. Res. 2, 012024 (2020).
[14] D. Yoon et al., Nano Lett.11, 3227 (2011).
[15] S. Linas et al., Phys. Rev. B 91, 075426 (2015).
[16] R. Yan et al., ACS Nano 8, 986 (2014).
[17] P. Kumar et al., App. Phys. Lett. 100, 222602 (2012).
[18] P. Kumar et al., J. Phys.: Condens. Matter 26, 305403 (2014).
[19] I. Bilgin et al., ACS Nano 12, 740 (2018).

[20]	B. Singh et al., J. Phys.: Condens. Matter 31, 065603 (2019).

[21]	P. Tonndorf et al., Opt. Exp. 21, 4908 (2013).

[22]	K. Kim et al., ACS Nano 10, 8113 (2016).

[23]	X. Lu et al., Adv. Mater. 27, 4502 (2015).

[24]	P. Soubelet et al., Phys. Rev. B 93, 155407 (2016).

[25]	B. Singh et al., Phys. Rev. Res. 2, 023162 (2020).

[26]	D. Kumar et al., J. Phys.: Condens. Matter 32, 415702 (2020).

[27]	Light Scattering in Solid II, edited by M. Cardona and G. Guntherodt, Springer Verlag Berlin (1982).

[28]	R. Loudon, Adv. Phys. 50, 813 (2001).

[29]	P.G. Klemens, Phys. Rev. 148, 845(1966).

[30]	J. Menendez and M. Cardona, Phys. Rev. B 29, 2051 (1984).

[31]	S. H. El-Mahalaway and B. L. Evans, Appl. Cryst. 9, 403 (1976).

[32]	Y. Ding and B. Xiao, RSC Adv. 5, 18391 (2015).

[33]	C. Sevik, Phys. Rev. B 89, 035422 (2014).

[34]	Standard Reference Material 739 Certificate; National Institute of Standards and Technology, 1991.

[35]	X. Huang et al., Sci. Rep. 6, 32236 (2016).

[36]	M. Yang et al., App. Phy. Lett. 110, 093108 (2017).

[37]	X. Hu et al., Phys. Rev. Lett. 120, 055902 (2018).

[38}	L. Su et al., Nanoscale 6, 4920 (2014).

[39]	C. K. Gan and Y. Y. F. Liu, Phys. Rev. B 94, 134303 (2016).

[40]	D. Kumar et al., J. Phys.: Condens Matter 31, 505403 (2019).

[41]	R. Fivaz and E. Mooser, Phys. Rev. 163, 743 (1967).

[42]	N. Bonin et al., Phys. Rev. Lett. 99, 176802 (2007).

[43]	C. Cong and T. Yu, Nat. Commun. 5, 4709 (2014).

[44]	D. Kumar et al., Nanotechnology. 32, 285705 (2021).

[45]	R. M. Martin and T.C. Damen, Phys. Rev. Lett. 26, 86 (1971).

[46]	R.M. Martin and C.M. Varma, Phys. Rev. Lett. 26, 1241 (1971).

[47]	H.G. Park et al., Sci. Rep. 8, 3173 (2018).

[48]	D.V. Tuan et al., Phys. Rev. B 98, 125308 (2018).

**Table-I:** List of the experimentally observed modes along with their symmetry assignments and frequency at 4 K for the 1L and 2L MoSe$_2$. Units are in cm$^{-1}$.

| Mode assignment | Frequency ($\omega$) | |
| --- | --- | --- |
| | 1L | 2L |
| S1 [$E^2_{2g}(\Gamma)$] | - | 17.8±0.02 |
| S2 [$TA(M-K)$] | 124.8±0.8 | 121.9±0.3 |
| S3 [$ZA(M-K)$] | 129.7±0.5 | 127.6±0.4 |
| S4 [$E^1_{2g} - LA(M)$] | 140.7±0.8 | 137.8±0.3 |
| S5 [$LA(M)$] | 152.1±0.3 | 147.3±0.3 |
| #1 | 160±1.0 | - |
| S6 [$E_{1g}(\Gamma)$] | 174.1±0.6 | 168.8±0.4 |
| S7 [$A_{1g}(\Gamma)$] | 242.3±0.1 | 242.2±0.02 |
| S8 [$E^1_{2g}(\Gamma)$] | 287.9±0.2 | 285.9±0.2 |
| #2 | 298.0±0.5 | |
| S9 [$2LA(M)$] | 305.5±0.3 | 302.8±0.4 |
| #3 | 312.7±0.8 | - |
| S10 | 321.2±0.1 | 318.9±0.2 |
| S11 | - | 342.5±0.5 |
| S12 [$A^2_{2u}(\Gamma)$] | 355.1±0.4 | 354.0±0.03 |
| S13 [$A_{1g}(M) + LA(M)$] | 364.3±1.0 | 362.8±0.7 |
| #4 | 387.4±2.5 | 389.2±2.5 |
| S14 [$TA(M) + 2LA(M)$] | 414.6±1.2 | 411.4±0.6 |
| S15 [$E^1_{2g}(M) + LA(M)$] | 432.7±0.3 | 430.2±0.2 |
| S16 $A^2_{2u}(M) + LA(M)$ | 444.6±0.2 | 441.7±0.3 |
| S17 [$3LA(M)$] | 457.2±0.1 | 453.9±0.1 |
| S18 [$TA(M) + 3LA(M)$] | 569.7±0.4 | 567.0±0.4 |
| S19 [$E^1_{2g}(M) + 2LA(M)$] | 584.6±0.2 | 581.6±0.1 |
| S20 [$4LA(M)$] | 598.2±0.1 | 595.4±0.4 |

**Table-II:** Room temperature linear and volumetric thermal expansion coefficients (TEC) for 1L and 2L MoSe$_2$ extracted using $E_{2g}^1$ and $A_{1g}$ modes. Units are in 10$^{-6}$ K$^{-1}$.

| MoSe$_2$ | $\alpha(E_{2g}^1)$ | $\alpha(A_{1g})$ | $\alpha_v = 2\alpha_a + \alpha_c$ |
|---|---|---|---|
| 1L | 4.1 | 13.2 | 21.4 |
| 2L | 9.1 | 5.3 | 23.5 |

**Table-III:** List of the fitting parameters obtained from FWHM of the first-order optical phonon modes for 1L and 2L MoSe$_2$. Units are in cm$^{-1}$.

| Modes | Electron-phonon + phonon-phonon coupling model | | | | | |
|---|---|---|---|---|---|---|
| | 1L | | | 2L | | |
| | C | $\gamma_{ph-ph}(0)$ | $\gamma_{e-ph}(0)$ | C | $\gamma_{ph-ph}(0)$ | $\gamma_{e-ph}(0)$ |
| $E_{2g}^2(\Gamma)$ | - | - | - | 0.007±0.001 | 1.1±0.04 | 1.3±0.3 |
| $E_{1g}(\Gamma)$ | 1.1±0.3 | 3.5±1.5 | 0.06±1.5 | 0.9±0.2 | 2.8±0.9 | 1.0±0.9 |
| $A_{1g}(\Gamma)$ | 0.2±0.1 | 1.4±0.02 | 0.02±0.01 | 0.3±0.02 | 0.6±0.1 | 0.7±0.08 |
| $E_{2g}^1(\Gamma)$ | - | - | - | - | - | - |
| $A_{2u}^2(\Gamma)$ | - | - | - | 0.5±0.3 | 1.9±0.8 | 0.1±0.6 |

**FIGURES:**

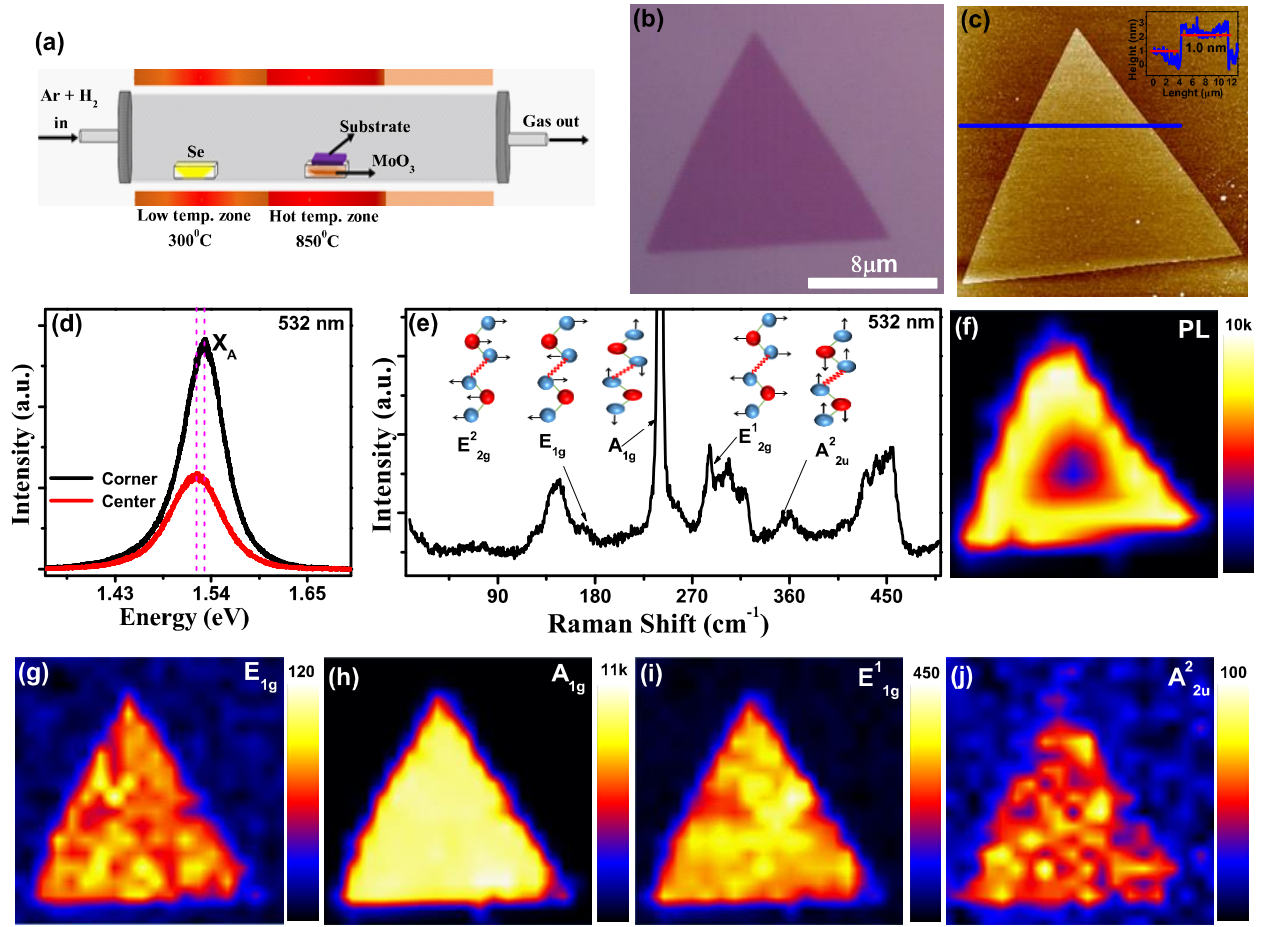

**FIGURE 1:** (a) Schematic diagram of CVD growth MoSe$_2$. (b) Optical and (c) AFM image of the triangular 1L MoSe$_2$ flake. (d) PL spectrum collected from the corner and center of MoSe$_2$ flake (e) Room temperature Raman spectrum. Insets show the ball stick model representing the vibrations of the modes. (f) PL intensity mapping image corresponding to the PL emission excitonic band ($A_{ex}$). (g-j) Raman intensity mapping images of the $E_{1g}$, $A_{1g}$, $E^1_{2g}$ and $A^2_{2u}$ phonon modes.

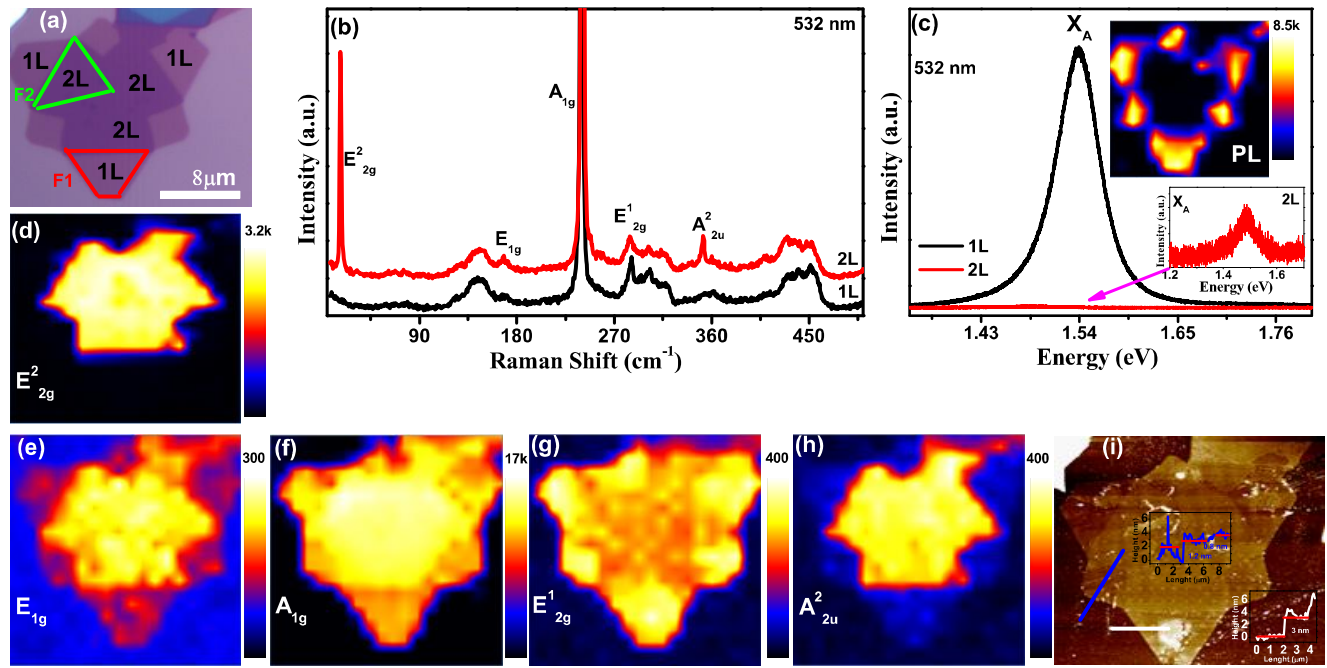

**FIGURE 2:** (a) Optical micrograph of the 1L and the 2L thickness regions of the MoSe$_2$ flake. The areas of focus are indicated by F1 and F2. Room-temperature (b) Raman and (c) PL spectrum, collected from the F1 (black) and F2 (red). The inset spectrum shows the PL peak position collected from the F2 and, the inset image illustrates the PL intensity mapping image of the $A_{ex}$ excitonic band. (d-h) Raman intensity map of the $E_{2g}^2$, $E_{1g}$, $A_{1g}$, $E_{2g}^1$ and $A_{2u}^2$ modes. (i) AFM image.

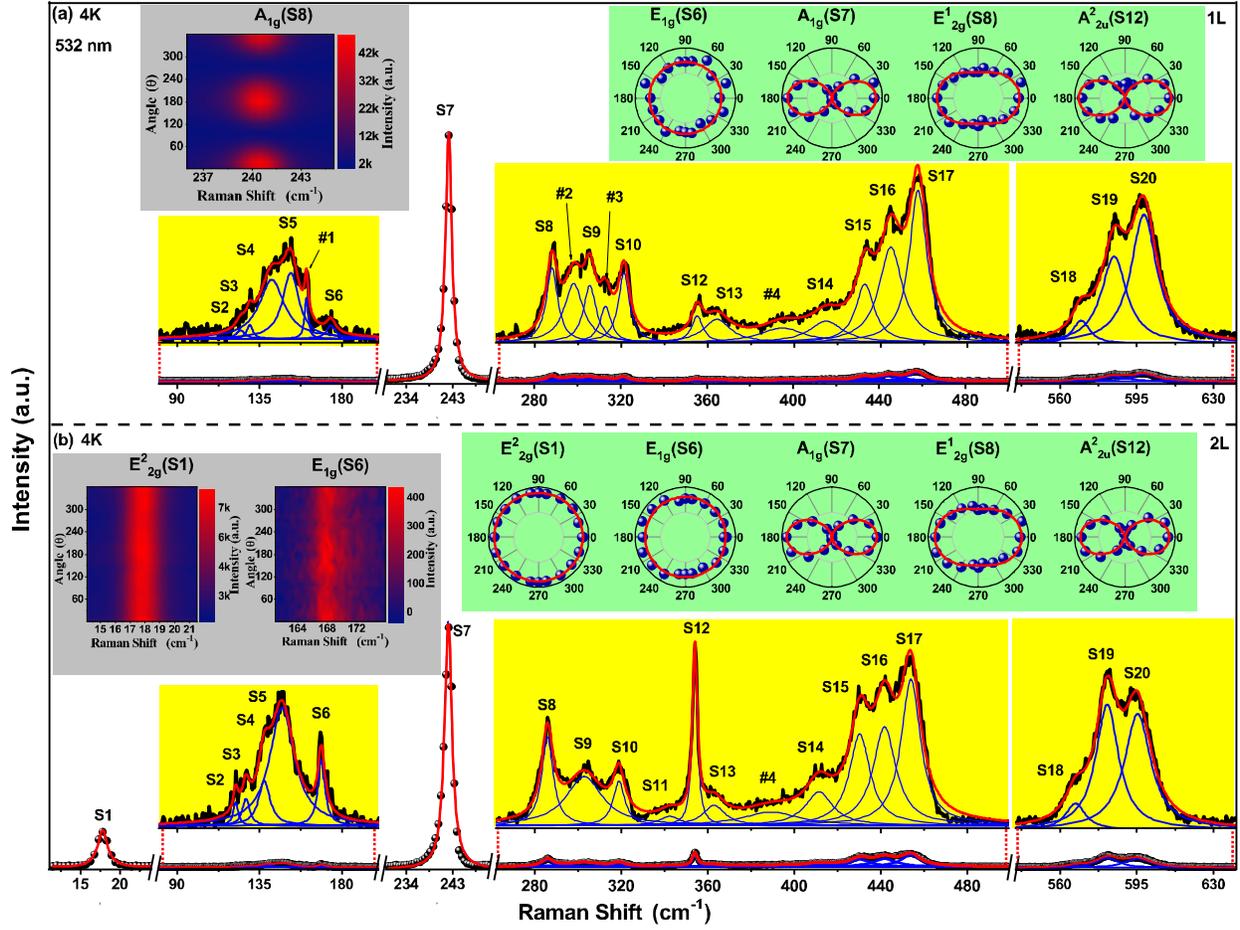

**FIGURE 3:** Raman spectra of 1L (a) and 2L (b) MoSe$_2$ at 4 K. Insets in the yellow shaded area show the amplified spectra. Insets in the green shaded area show the intensity polar plot of the $E^2_{2g}$, $E_{1g}$, $A_{1g}$, $E^1_{2g}$ and $A^2_{2u}$ modes; solid red lines are the fitted curves as described in the text. Insets in grey colour area illustrate the 2D colour contour maps of the Raman intensity versus Raman shift and as a function of polarisation angle for 1L and 2L MoSe$_2$.

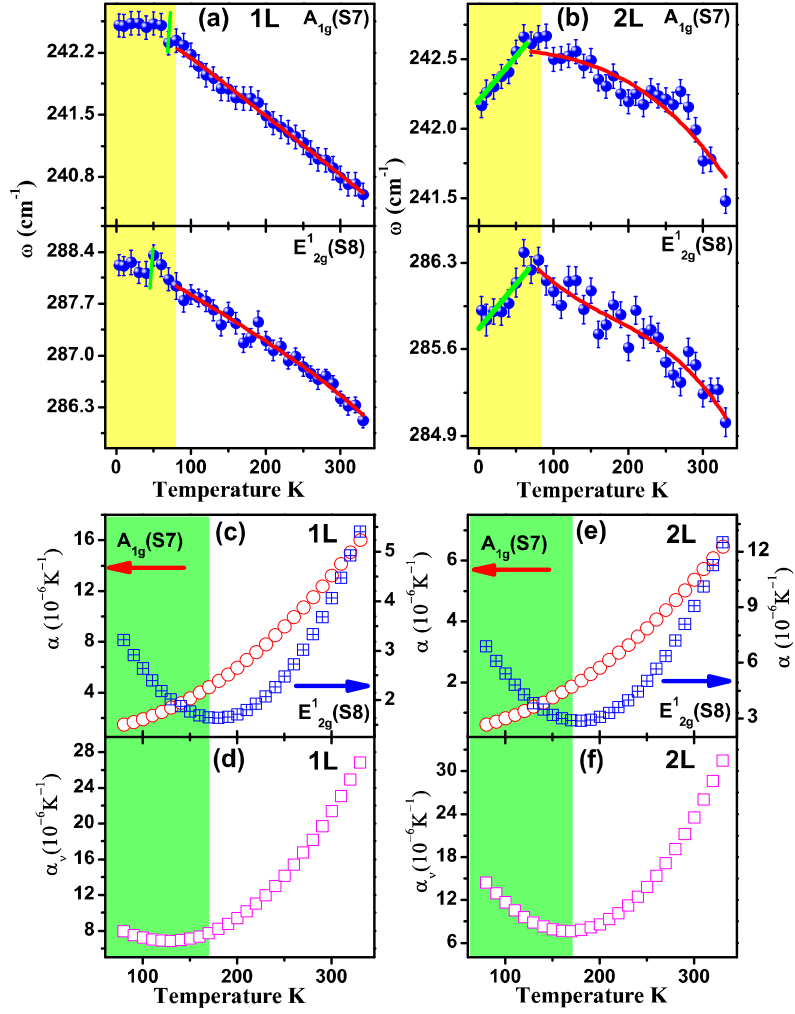

**FIGURE 4:** (a) and (b) Temperature dependence of the frequency of $A_{1g}$ and $E^1_{2g}$ modes for 1L and 2L MoSe$_2$, respectively. Solid red lines are the fitted curves as described in the text, and the solid green lines are a guide to the eye. The shaded part illustrates the region where the mode frequency shows anomalous behaviour. (c) and (e) Linear TEC corresponding to the $A_{1g}$ (red) and $E^1_{2g}$ (blue) modes and (d) and (f) Volumetric TEC in the temperature range of 80-330 K for 1L and 2L MoSe$_2$, respectively.

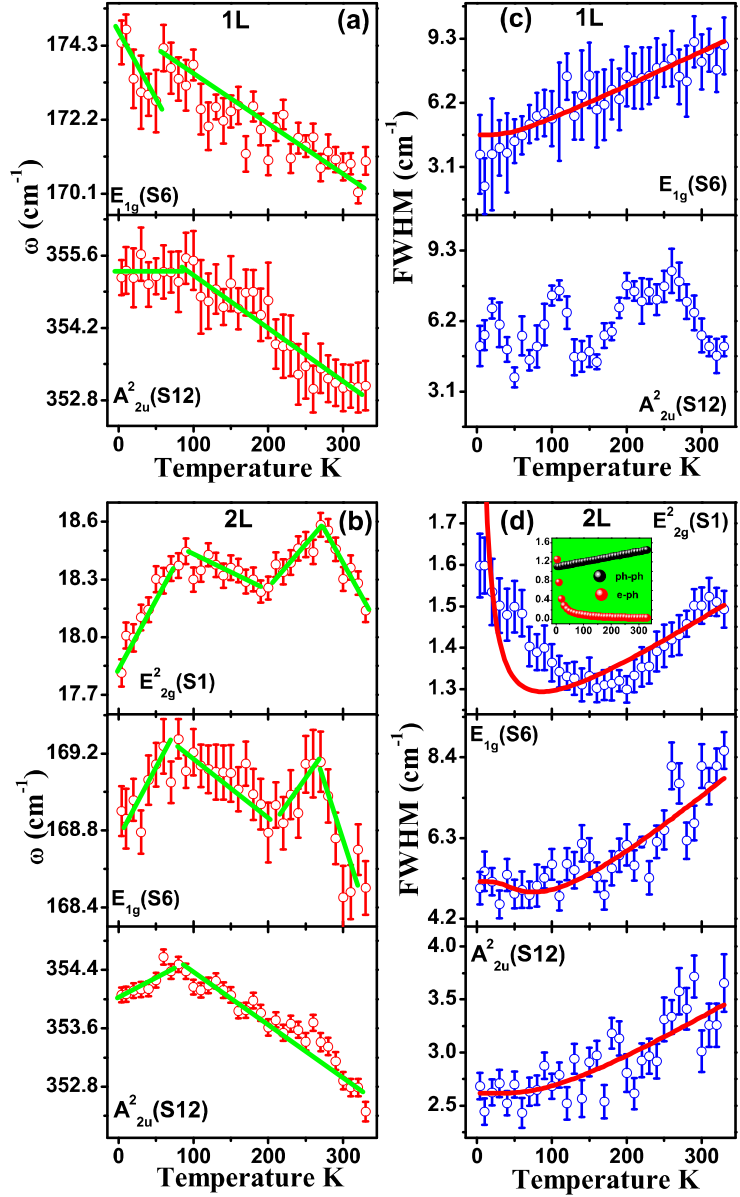

**FIGURE 5:** Temperature dependence (a) frequency, and (c) FWHM of $E_{1g}$ and $A_{2u}^2$ modes for 1L. Temperature dependence (b) frequency, and (d) FWHM of $E_{2g}^2$, $E_{1g}$ and $A_{2u}^2$ modes for 2L. Solid red lines are the fitted curves described in the text, and solid green lines are guide to the eye. Inset black and red plots in (d) describe individual contributions to the linewidth from phonon-phonon and electron-phonon coupling, respectively.

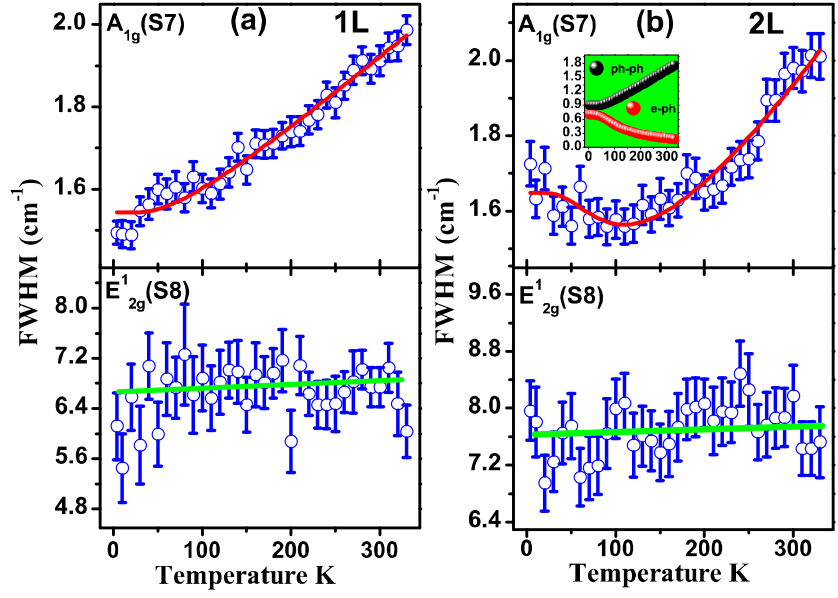

**FIGURE 6:** Temperature dependence FWHM of $A_{1g}$ and $E^1_{2g}$ modes for (a) 1L; (b) 2L. Solid red lines are the fitted curves described in the text, and solid green lines are guide to the eye. Inset black and red plots in (b) describe individual contributions to the linewidth from phonon-phonon and electron-phonon coupling, respectively.

# Supplementary Information:

**Electron-Phonon Coupling, Thermal Expansion Coefficient, Resonance Effect and Phonon Dynamics in High Quality CVD Grown Mono and Bilayer MoSe$_2$**

Deepu Kumar[1#], Vivek Kumar[1], Rahul Kumar[2], Mahesh Kumar[2], Pradeep Kumar[1*]

[1]*School of Basic Sciences, Indian Institute of Technology Mandi, 175005, India*
[2]*Department of Electrical Engineering, Indian Institute of Technology Jodhpur, 342001, India*

#E-mail: deepu7727@gmail.com
*E-mail: pkumar@iitmandi.ac.in

## 1. Synthesis of MoSe$_2$ using CVD

The triangular-shaped 2D MoSe$_2$ was synthesized using a three zones thermal chemical vapour deposition (CVD) system equipped with a quartz tube (2-inch diameter) under atmospheric pressure. Figure 1 (a) (see in main text) is illustrating a schematic diagram of the CVD system and showing that the two solid precursors, selenium (Se) and molybdenum trioxide (MoO$_3$), were loaded in the first and second zone of the CVD system, respectively. In this typical synthesis, 1.0 g of Se powder (99.9%, Sigma-Aldrich) and 0.04 g of MoO$_3$ powder (99.9%, Sigma-Aldrich) were placed 15 cm away from each other, and a cleaned SiO$_2$/Si substrate was put face down on MoO$_3$. Before the chemical reaction, the tube was evacuated by a rotary pump and purged several times with ultra-high pure Ar gas to remove oxygen and other contamination. After achieving atmospheric pressure with purging Ar gas, the temperature of the MoO$_3$ in the downstream zone was increased up to 850°C with a heating rate of 20°C per min and the Se in the upstream zone was heated at 300°C. The chemical reaction was performed for 30 min at above temperature values in the presence of carrier gases' mixture (1% H$_2$ + 99% Ar) with a flow of 60 SCCM under atmospheric pressure. Finally, the furnaces of the zones were switched off, and the tube was cooled naturally back to room temperature.

## 2. Measurement techniques and Raman Phonon modes

The sample quality and crystal structure were assessed using Raman and PL features and the intensity mapping at room temperature. Raman and photoluminescence (PL) measurements were performed with Horiba LabRAM HR evolution in backscattering geometry. A 532 nm (2.33 eV) laser was used to excite the Raman and PL spectra. Laser power was set to be very low ~ 0.2 mW to avoid any local heating, as well as damage to the sample. The laser beam was focused on the sample using a 100x long working objective lens, and the same objective lens was used to collect scattered light from the sample. The scattered light from the sample was detected by using 1800 and 600 grating coupled with Peltier cooled Charge Coupled Device (CCD) detector for Raman and PL measurements, respectively. For polarized Raman scattering, a polarizer and an analyzer have been inserted into the incident light path on the sample and scattered light from the sample, respectively. The polarization-dependent Raman measurements have been performed at room temperature with rotating the direction of incident light with an interval of $20^0$, while keeping fixed the direction of scattered light and position of the sample. For temperature-dependent Raman measurements, a 50x long working objective lens was used both to focus the incident beam on the sample and to collect the scattered light. The temperature-dependent Raman measurements were carried out using a closed cycle refrigerator (Montana Cryostat) in a wide temperature range from 4 to 330 K, with an interval of 10 K and temperature accuracy of ± 0.1 K and waiting time for each Raman measurement was ~ 10 minutes for better temperature stability.

Bulk MoSe$_2$ belongs to the point group $D_{6h}^4$ ( space group $P6_3/mmc, \#194$), the unit cell is composed of two atomic formula (Z=2) with 6 atoms, which results in 18 phonon branches at the $\Gamma$ point of the BZ, and these phonon branches can be expressed by following irreducible representation as $A_{1g} + 2A_{2u} + 2B_{2g} + B_{1u} + E_{1g} + 2E_{1u} + 2E_{2g} + E_{2u}$ [1]. As the thickness of the material decreases from bulk to few layers, resulting in symmetry variations due to the loss of the translational symmetry perpendicular to the basal plane. For example, monolayer exhibiting the non-centrosymmetric, and belonging to the point group $D_{3h}^1$ (space group $P\bar{6}m2, \#187$), there are 3 atoms per unit cell, giving into 9 phonon branches with the irreducible representation $A_1' + 2E' + 2A_2'' + E''$. The bilayer exhibits the centrosymmetric nature belonging to the point group $D_{3d}^3$ (space group $P\bar{3}m1, \#164$)). There are 6 atoms per unit cell, giving 18 phonon branches with an irreducible representation $3A_{1g} + 3E_g + 3A_{2u} + 3E_u$. The phonon modes with symmetry $E$ are the in-plane vibration of atoms and are doubly degenerate, while the phonon modes with symmetry $A$ and $B$ correspond to the non-degenerate out-of-plane vibrations of atoms.

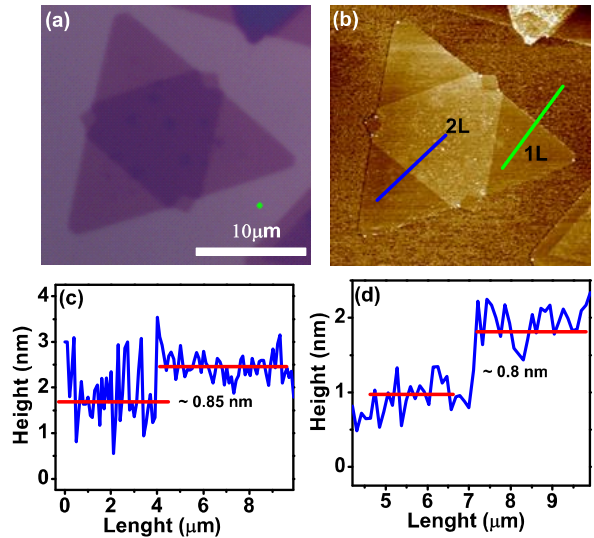

**FIGURE S1:** (a) Optical micrograph of CVD grown MoSe$_2$ flake on SiO$_2$/Si substrate consisting of the monolayer (1L) and bilayer (2L) thickness regions (c) Atomic Force Microscope (AFM) image. (c) Height profile taken along the green line for 1L from the substrate. (d) Height profile taken along the blue line for 2L from 1L.

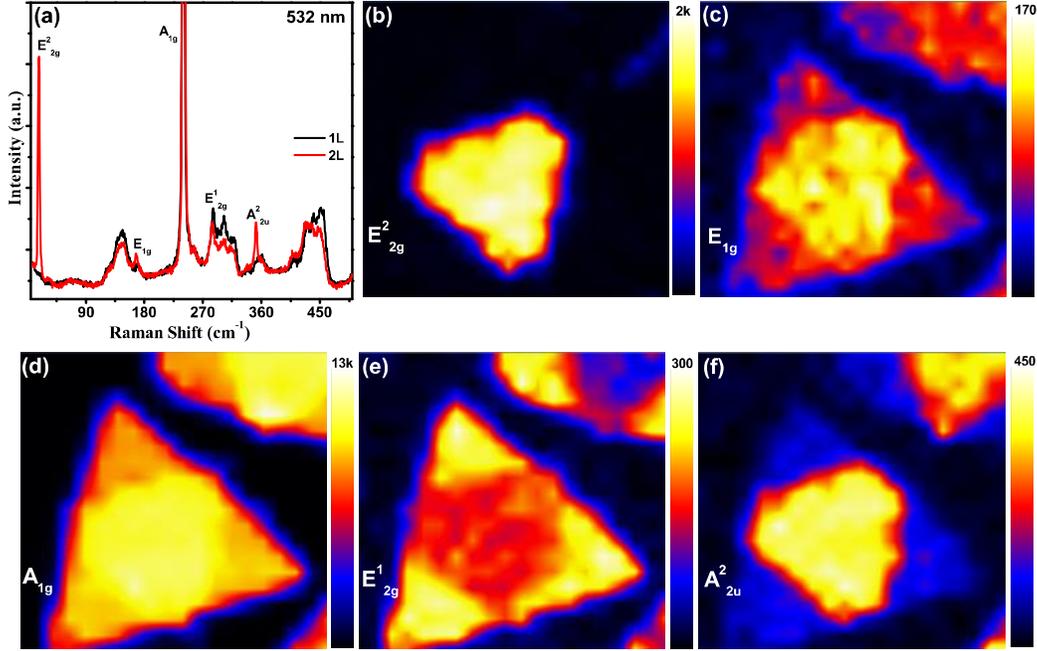

**FIGURE S2:** (a) Room-temperature Raman spectrum in the spectral range of 5-500 cm$^{-1}$ collected from 1L(black) 2L (red) by using a 532 nm laser. (b-f) Raman intensity mapping images of the $E_{2g}^2$, $E_{1g}$, $A_{1g}$, $E_{2g}^1$ and $A_{2u}^2$ phonon modes, respectively.

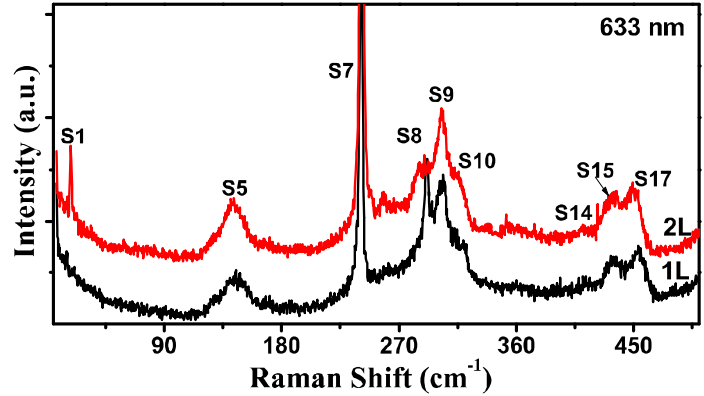

**FIGURE S3** Room-temperature Raman spectrum in the spectral range of 5-500 cm$^{-1}$ collected from 1L (black) 2L (red) by using a 632.8 nm laser.

## 3. Temperature dependence of the first-order acoustic and second and higher-order modes

Figure S4 (a) and S4 (b) Temperature evaluation of the Raman spectra in the temperature range of 4-330 K for 1L and 2L MoSe$_2$, respectively. Figures S5 (a) and S5 (b) illustrate the temperature dependence of the mode frequency ($\omega$) and FWHM of the S3 and S5 phonon modes for the monolayer and bilayer MoSe$_2$, respectively. Following observations can be made from our temperature-dependent Raman measurements: (i) For monolayer, frequency of both modes show normal behaviour i.e. mode hardening with decreasing temperature; on the other hand, for bilayer, we observe the anomalous behaviour of both modes i.e. mode softening with decreasing temperature. (ii) For monolayer, FWHM of S3 mode show normal behaviour, i.e. FWHM decreases with decreasing in temperature, while FWHM of S5 mode show normal behaviour down to ~250 K with decreasing temperature, and below 200 K it exhibits nearly temperature-independent behaviour till 4 K. For bilayer, FWHM of the S3 (S5) modes show normal (anomalous) temperature-dependent behavior.

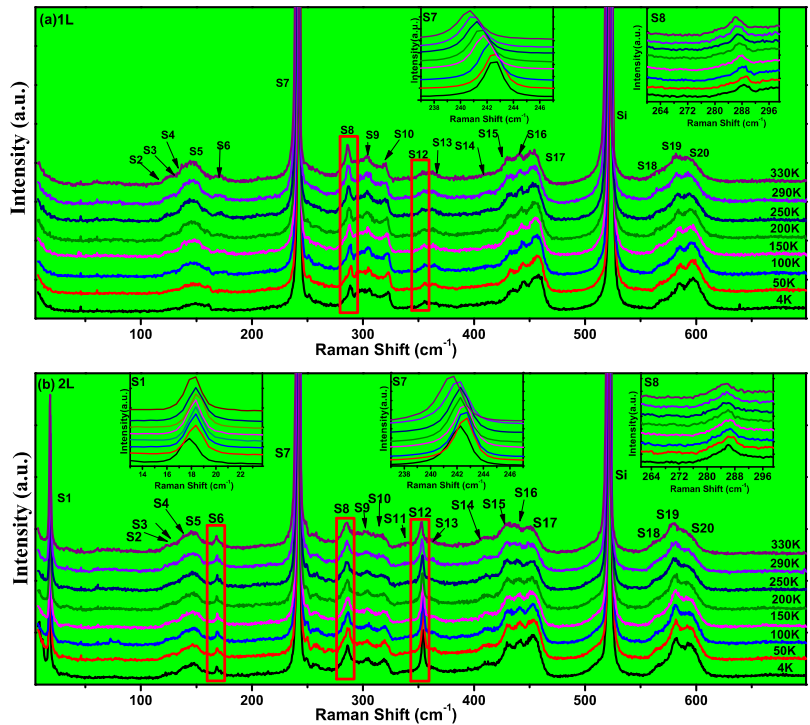

**FIGURE S4:** (a) and (b) Temperature evaluation of the Raman spectra in the temperature range of 4-330 K for 1L and 2L MoSe$_2$, respectively. Insets (a) show the expanded region of the $A_{1g}$ (S7) and $E^1_{2g}$ (S8) modes for monolayer; and insets (b) show the expanded region of the $E^2_{2g}$ (S1), $A_{1g}$ (S7) and $E^1_{2g}$ (S8) modes for bilayer.

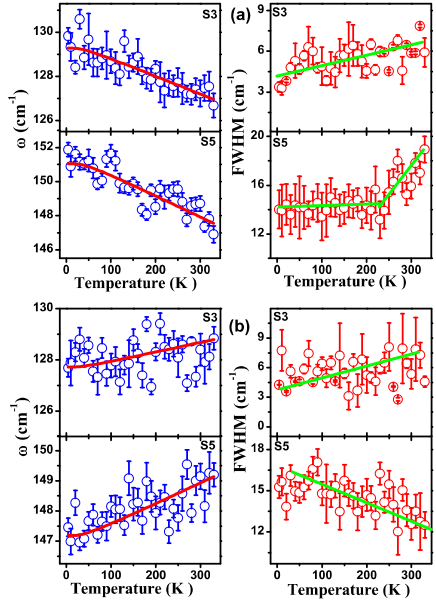

**FIGURE S5:** Temperature dependence of the frequency and FWHM of the S3 and S5 modes for (a) 1L (b) 2L. Solid red lines are the fitted curves via a three-phonon anharmonic model described in the text, and solid green lines are guide to the eye.

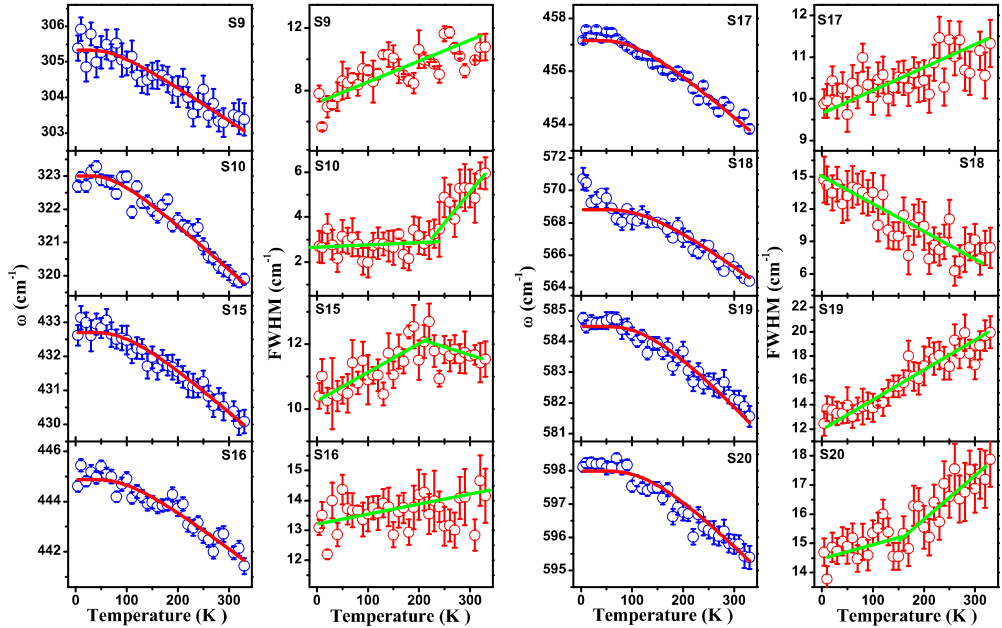

**FIGURE S6:** Temperature dependence of the phonon mode frequencies and FWHM of the modes S9, S10, S15, S16 S17, S18, S19 and S20 for 1L $MoSe_2$. Solid red lines are the fitted curves via three phonon anharmonic model as described in the text and solid green lines are a guide to the eye.

Figure S6 illustrates the temperature dependence of mode frequency ($\omega$) and FWHM of S9, S10, S15, S16 S17, S18, S19 and S20 phonon modes for the monolayer. Following observations can be made: (i) Frequency of all phonon modes S9, S10, S15, S16 S17, S18, S19 and S20 show normal temperature dependence. (ii) FWHM of the modes S9, S16, S17, S19, S20 show normal temperature dependence. FWHM of S10 modes show normal behaviour only down to ~250 K, and below 200 K, it remains nearly temperature-independent down to the lowest recorded temperature (4 K). FWHM of S15 mode increases slightly with decreasing temperature till ~ 200 K; on further cooling, it starts to decrease. FWHM of S18 mode increases with decreasing temperature from 330 K to lowest recorded temperature, 4 K. The temperature dependence of the mode frequency ($\omega$) and FWHM of S9, S10, S15, S16 S17, S18, S19 and S20 phonon modes for the bilayer is shown in Fig. S7. Following observations can be made: (i) Frequency of all phonon modes S9, S10, S15, S16 S17, S18, S19 and S20 except S9 show normal temperature dependence. (ii) FWHM of the modes S9, S10, S15, S16, S17, S19 show normal temperature dependence. FWHM of S18 mode increases with decreasing temperature from 330 K to lowest recorded temperature, 4 K, FWHM of S20 mode first decreases with decreasing temperature down to ~100 K, on further cooling, it starts to increase. The solid red lines in Fig. S5, Fig.S6 and Fig. S7 are the fitted curves via three phonon anharmonic model as described in the main text, and the fitting is in good agreement with the experimental data above ~100 K, and the solid green lines are guide to the eye.

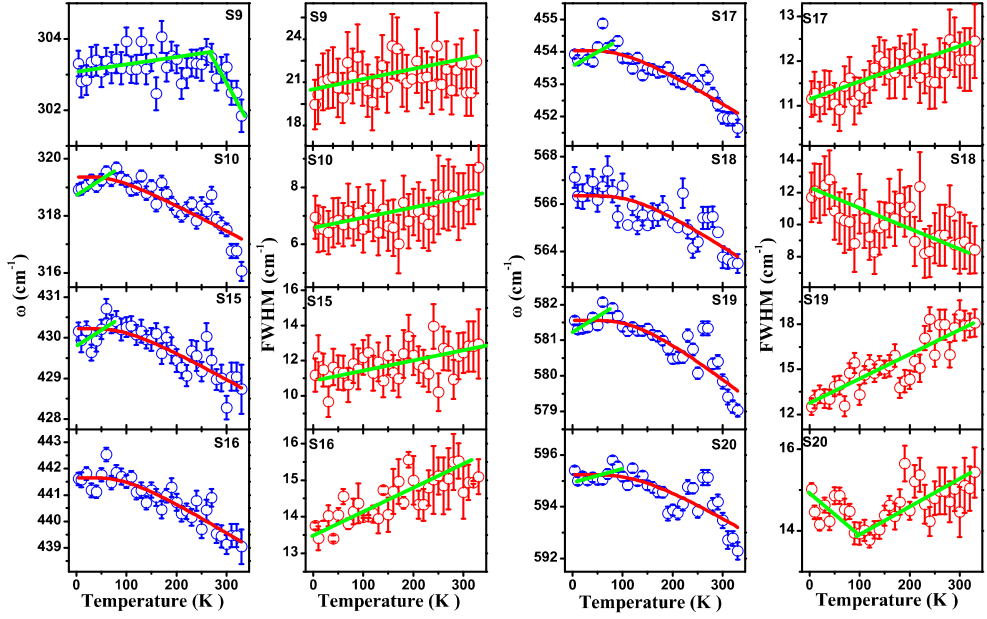

**FIGURE S7:** Temperature dependence of the phonon mode frequencies and FWHM of the modes (a) S9, S10, S15, and S16; (b) S17, S18, S19 and S20 for bilayer MoSe$_2$. Solid red lines are the fitted curves via three phonon anharmonic model as described in the main text and solid green lines are guide to the eye.

## 4. Temperature-dependent Raman intensity of the phonon modes

In most of the previous literature, we note that the Raman intensity of the phonons and its dependence on the temperature has not been touched upon. While, in these 2D materials, the temperature-dependent Raman intensity of the phonon modes may provide rich information about the electronic and optical properties of the materials as well as tuning of these properties as a function of temperature. Very few reports for these 2D materials are available to the best of our knowledge, wherein Raman intensity of the phonon modes and its dependence on the temperature has been studied in detailed [2-3]. However, no such reports are available for the case of MoSe$_2$. This section focuses on the Raman intensity of the phonon modes and its temperature dependence for mono and bilayer. From Fig. S4 (a) and S4 (b), one could see that S6 ($E_{1g}$) mode is prominent at high temperature in monolayer as well as bilayer, while S8 ($E_{2g}^1$) mode is found to be intense at high temperature only for monolayer and it becomes intense at low temperature for the bilayer. Temperature-dependent intensity evaluation of the S12 ($A_{2u}^2$) mode is not very clear for the case of the monolayer, maybe because of the very weak signal, while for the bilayer, one could see that it become intense at low temperature compared to that of high temperature. To quantitatively understand the temperature-dependent Raman intensity of the phonon modes, we extracted the intensity of the individual phonons by using the Lorentzian function. Figure S8 (a) and S8 (b) show the integrated intensity of the few prominent modes such as S5 ($LA$), S6 ($E_{1g}$), S7 ($A_{1g}$), S8 ($E_{2g}^1$) and S12 ($A_{2u}^2$) as a function of temperature for monolayer and bilayer, respectively. For the monolayer, we observed that the intensity of the S5 ($LA$), S6 ($E_{1g}$), S8 ($E_{2g}^1$) and S12 ($A_{2u}^2$) modes decrease with decreasing temperature, while it remains temperature independent for the S7 ($A_{1g}$) mode. For bilayer, the intensity of the S5 ($LA$), S6 ($E_{1g}$), S8 ($E_{2g}^1$) and S12 ($A_{2u}^2$) modes increases, while it decreases for the S6 ($E_{1g}$), S7 ($A_{1g}$) modes, with decreasing temperature. Figure S8 (c) shows the intensity ratio of the modes as a function of temperature for monolayer and bilayer. We observed that the intensity ratio of the $A_{1g}$ mode with respect to $E_{2g}^1$ mode increases (decreases) with decreasing temperature for monolayer (bilayer), while the intensity of the $A_{2u}^2$ mode with respect to $E_{1g}$ mode increases for both layers. The intensity ratio of the $2LA$ mode with respect to $LA$ increases (decreases) with the decrease in temperature for the monolayer (bilayer), while an increase is observed in the intensity ratio of the $3LA$ and $4LA$ modes with respect to $LA$ mode with decreasing temperature in both these systems. At low temperature, $3LA$ and $4LA$ modes are ~ 4-5 (~ 2) times stronger compared to that of $LA$ mode for the monolayer (bilayer). $A_{2u}^2$ mode is ~ 4-5 times stronger than $E_{1g}$ in both monolayer and bilayer at low temperature. On the other hand, $A_{1g}$ mode is ~14-15 times stronger than that of $E_{2g}^1$ mode at low temperature for the monolayer, while it is ~12-13 times stronger than that of $E_{2g}^1$ mode at high temperature for the bilayer.

Now we focus on understanding the temperature-dependent variations in the intensity of the phonon modes. In these kinds of 2D materials, the variations in the intensity of the phonon modes as a function of temperature preliminary may be understood considering the tuning of resonance conditions with temperature variations. The tuning of the resonance conditions as a function of incident photon energies and its impact on the intensity of the phonon modes in these TMDCs materials, including MoSe2, has been studied in several earlier studies [4-5]. Phonon modes $A_{1g}$, $E_{2g}^1$ and $LA$ and their overtone become prominent when excitation energies resonate with the $A$ and $B$ excitonic energy band. In contrast, the intensity of the $E_{1g}$ and $A_{2u}^2$ modes is enhanced when laser energy resonates with the C exciton energy. In the quantum mechanical picture, the Raman scattering intensity of phonon mode may be given as [3, 6]

$$I \propto \left| \frac{1}{[E(T) - E_L + i\gamma(T)][E(T) - E_s + i\gamma(T)]} \right|^2 \qquad (1)$$

where $E(T)$ and $\gamma(T)$ are the temperature-dependent transition energies and damping constants of excitonic bands, respectively, while $E_L$ ($E_S$) is the energy of the incident (scattered) photons, which is fixed in our case. The energy of the excitons are strongly dependent on temperature, and it could be tuned as a function of temperature. Therefore, in the present case as well, the resonance conditions may be changed by changing the temperature of the sample. The energy of the excitonic bands increases with descreasing temperature [7]. As the temperature increases, the energy of C exciton move away from the incident laser photon energy, while at the same time the energies of A or B excitons approachs the incident laser photon energy.

Therefore, the intensity of the modes $A_{1g}$, $E_{2g}^1$ and $LA$ as well as its overtone, should be expected to increase with decreasing temperature, while the opposite nature is anticipated for the $E_{1g}$ and $A_{2u}^2$ modes. The temperature-dependent intensity of $E_{1g}$ mode for both layers is in good agreement with the above statement, but $A_{2u}^2$ mode only shows good agreement for monolayer, while it violates for bilayer, see Fig. S8 (a) and S8 (b). The intensity of the $A_{1g}$, $E_{2g}^1$ and $LA$ also shows different temperature dependence for different thickness. For example $A_{1g}$ mode is intense in bilayer compared to monolayer, while opposite nature is observed for the $E_{2g}^1$ mode (see Fig. 3 in main text). Therefore, the observed different temperature-dependent intensity for different modes and different layer thickness could not be explained, considering only the tuning of the resonance condition. It is reported that for these 2D materials, the wave function of A and B exciton are strongly confined to the single individual layer, with only a slight overlap to the adjoining layers [8]. On the other hand, the wave function of the *C* exciton extended over the entire thickness in the layer system. Moreover, the energy of *A* and *C* excitons is observed to depend strongly on the number of layers, while the energy of B exciton shows very weak dependence. The spatial confinement of the wave function of the excitons and layers dependence energies of the excitons possesses different resonance effect for different phonon modes. This may give the different temperature and thickness dependence of the intensity of the phonon modes. Further, TEC and TEC mismatch between $MoSe_2$ and substrate also significantly impact the intensity of the phonon modes. Because these effects are strongly dependent on the layer thickness and temperature, and these may be other possible reasons for different temperature and different thickness-dependent intensity of the observed phonons modes in $MoSe_2$ system studied here.

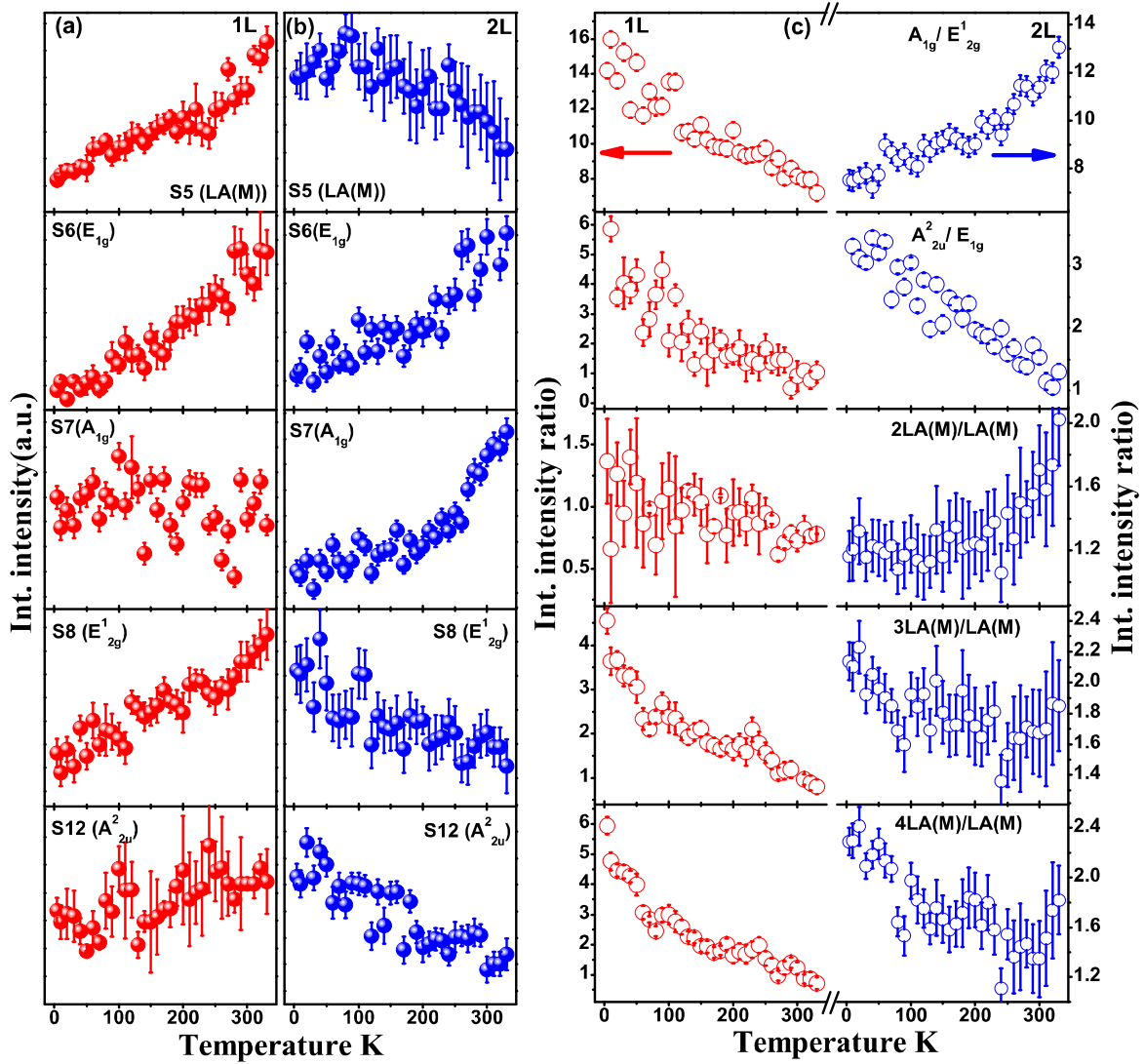

**FIGURE S8:** (Color online) (a) and (b) Temperature dependence of the integrated intensity of S5 ($LA$), S6 ($E_{1g}$), S7 ($A_{1g}$), S8 ($E^1_{2g}$) and S12 ($A^2_{2u}$) phonon modes for monolayer and bilayer, respectively. (c) Temperature dependence of the integrated intensity ratio of the phonon modes.

**References:**

[1]   N. Scheuschner et al., Phys. Rev. B 91, 235409 (2015).

[2]   D. Kumar et al., Nanotechnology. 32, 285705 (2021).

[3]   H. Zobeiri et al., Nanoscale 12, 6064 (2020).

[4]   K. Kim et al., ACS Nano 10, 8113 (2016).

[5]   P. Soubelet et al., Phys. Rev. B 93, 155407 (2016).

[6]   Light Scattering in Solid II, edited by M. Cardona and G. Guntherodt, Springer Verlag Berlin (1982).

[7]   H.G. Park et al., Sci. Rep. 8, 3173 (2018).

[8]   G.Y. Jia et al., Mater. Chem. C 4, 8822 (2016).